\documentclass[a4paper,12pt]{article}

\usepackage{amssymb,amsmath,cite,empheq,epsfig,graphicx,psfrag}

\newcommand{\A}{A}                        % action
\newcommand{\Aeik}{\A_{\mathrm{eik}}}     % eikonal action
\newcommand{\As}{\mathcal{A}}             % A storto = action / (2\pi R\sqrt{s})
\newcommand{\amp}{{\cal M}}               % amplitude
\newcommand{\bs}{\boldsymbol}
\newcommand{\bt}{\boldsymbol{b}}
\newcommand{\Caption}[1]{\caption{\it #1}}% didascalia personalizzata
\newcommand{\cl}{\mathrm{cl}}             % classical
\newcommand{\delay}{\mathrm{delay}}       % delay
\newcommand{\deltap}{\delta_{+}}          % delta mass-shell
\newcommand{\dif}{\mathrm{d}}             % finite-dimensional differential
\newcommand{\dq}{\partial^2}              % (d/dz)^2
\newcommand{\dsq}{\partial^{*2}}          % (d/dz*)^2
\newcommand{\dmq}{|\partial|^2}           % |d/dz|^2
\newcommand{\eik}{\mathrm{eik}}          % eikonal
\newcommand{\e}{\varepsilon}
\newcommand{\esp}[1]{\mathrm{e}^{#1}}    % esponential
\newcommand{\fourth}{{\textstyle\frac{1}{4}}}
\newcommand{\gr}{\boldsymbol{\nabla}}     % 2-D gradient
\newcommand{\Hc}{\mathcal{H}}
\newcommand{\half}{{\textstyle\frac{1}{2}}}
\newcommand{\hti}{\tilde{h}}
\newcommand{\imp}{\Longrightarrow}
\renewcommand{\in}{{\mathrm{in}}}          % incoming
\newcommand{\J}{{\cal J}}                 % current
\newcommand{\Jr}{{J_\textsc{r}}}          % rescattering current
\newcommand{\kt}{\boldsymbol{k}}
\renewcommand{\L}{\mathcal{L}}             % lagrangian
\newcommand{\ord}[1]{\mathcal{O}\left(#1\right)}
\newcommand{\out}{{\mathrm{out}}}          % outgoing
\newcommand{\Qt}{\boldsymbol{Q}}
\newcommand{\qt}{\boldsymbol{q}}
\newcommand{\reg}{\mathrm{reg}}
\newcommand{\rung}{\mathcal{R}}           % rung
\newcommand{\sgn}{\epsilon}        % sign as function (epsilon)
\newcommand{\sphi}{\Sigma}              % support of Phi
\newcommand{\halftheta}{{\textstyle{\frac{\theta}{2}}}}
\newcommand{\ui}{\mathrm{i}}             % imaginary unit
\newcommand{\vb}{\vec{b}}                % 3-vector b
\newcommand{\vk}{\vec{k}}                % 3-vector k
\newcommand{\vp}{\vec{p}}                % 3-vector p
\newcommand{\vQ}{\vec{Q}}                % 3-vector Q
\newcommand{\vq}{\vec{q}}                % 3-vector q
\newcommand{\vx}{\vec{x}}                % 3-vector x
\newcommand{\xt}{\boldsymbol{x}}

\title{{\bf Rescattering corrections and self-consistent metric in Planckian scattering}}

\author{
   M.~Ciafaloni
   \footnote{Email: ciafaloni@fi.infn.it}
   \\[1ex]
   {\sl\small c/o Dipartimento di Fisica, Universit\`a di Firenze}\\
   {\sl\small Via Sansone 1, 50019 Sesto Fiorentino, Italy}\\[10mm]
   D.~Colferai
   \footnote{Email: colferai@fi.infn.it}
   \\
   {\sl\small Dipartimento di Fisica, Universit\`a di Firenze and INFN, Sezione di Firenze}\\
   {\sl\small Via Sansone 1, 50019 Sesto Fiorentino, Italy}
}

\date{}

\begin{document}
%@@@@@@@@@@@@@@@@@@@@@@@@@@@@@@@@@@@@@@@@@@@@@@@@@@@@@@@@@@@@@@@@@@@@@@@@@@@@@@@

\maketitle

\begin{abstract}
Starting from the ACV approach to transplanckian scattering, we present a
development of the reduced-action model in which the (improved) eikonal
representation is able to describe particles' motion at large scattering angle and,
furthermore, UV-safe (regular) rescattering solutions are found and incorporated
in the metric. The resulting particles' shock-waves undergo calculable
trajectory shifts and time delays during the scattering process --- which turns
out to be consistently described by both action and metric, up to relative order
$R^2/b^2$ in the gravitational radius over impact parameter expansion. Some
suggestions about the role and the (re)scattering properties of irregular
solutions --- not fully investigated here --- are also presented.
\end{abstract}

\vskip 1cm

%\begin{minipage}{0.9\textwidth}
%\begin{flushright}
%  Draft $ $Revision: cc5af72780de $ $ \\
%  $ $Date: 2014/09/01 07:52:06 $ $
%\end{flushright}
%\end{minipage}
\vskip 1cm

%%%%%%%%%%%%%%%%%%%%%%%%%%%%%%%%%%%%%%%%%%%%%%%%%%%%%%%%%%%%%%%%%%%%%%%%%%%%%%%%%
\section{Introduction\label{s:intro}}
%%%%%%%%%%%%%%%%%%%%%%%%%%%%%%%%%%%%%%%%%%%%%%%%%%%%%%%%%%%%%%%%%%%%%%%%%%%%%%%%%

Interest in the gravitational $S$-matrix at transplanckian
energies~\cite{ACV88,GrMe87,ACV90,ACV93} has revived in the past
few-years~\cite{ACV07,CC08,CC09}, when explicit solutions of the so-called
reduced-action model~\cite{ACV93} have been found~\cite{ACV07}. The model is a
much simplified version of the ACV eikonal approach~\cite{ACV88,ACV90} to
transplanckian scattering in string-gravity, and is valid in the regime in which
the gravitational radius $R\equiv2G\sqrt{s}$ is much larger than the string
length $\lambda_s\equiv\sqrt{\alpha'\hbar}$, so that string-effects are supposed
to be small.

The reduced-action model~(sec.~\ref{s:rcram}) was derived by justifying the
eikonal form of the $S$-matrix at impact parameter $b$ on the basis of string
dynamics and by then calculating the eikonal itself (of order
$\sim\frac{Gs}{\hbar}\gg1$) in the form of a 2-dimensional action, whose power
series in $R^2/b^2$ corresponds to an infinite sum of proper irreducible
diagrams (the ``multi-H'' diagrams~\cite{ACV90,ACV93}), evaluated in the
high-energy limit. The model admits a quantum generalization~\cite{CC08} of the
$S$-matrix in the form of a path-integral --- with definite boundary conditions
--- of the reduced-action exponential itself.

The main feature of the model and of its boundary conditions is the existence of
a critical impact parameter $b_c\sim R$ such that, for $b>b_c$ the $S$-matrix
matches the perturbative series and is unitary, while for $b<b_c$ the field
solutions are complex-valued and the elastic $S$-matrix is suppressed
exponentially. The suppression exponent is of order
$\frac{Gs}{\hbar}\sim\frac{R^2}{\lambda_P^2}$ ($\lambda_P$ being the Planck
length) or, if we wish, of the same order as the entropy of a black-hole of
radius $R$. From various arguments we believe that in the region in which
$b<b_c$ (that is, $b$ is smaller than the gravitational radius), a
classical gravitational collapse is taking place.

A key issue related to the collapse region is the possible existence of
information loss at quantum level, which in the ACV model shows up as lack of
$S$-matrix unitarity. According to~\cite{CC08} such unitarity loss is
mostly related to a restrictive boundary condition, which is required for the
solutions of the model be UV-safe. If such condition is relaxed, further
solutions show up, which could contribute to unitarity, but are irregular ---
i.e.,\ dominated by planckian distances, region in which the model itself is
inadequate.

The main motivation of the present paper is to improve and complete some
unsatisfactory aspects of the reduced-action model which might be crucial at
planckian distances, but show up already at distances of order $R\equiv 4GE$,
the gravitational radius.  The new model that we shall present here features two
improvements. Firstly, the eikonal representation itself is embedded in three
dimensions, so as to be able to describe the motion of the Breit-frame during
the scattering process. By comparison, the usual two-dimensional representation
is inadequate outside the regime of very small angle scattering, and in
particular in the collapse region. In fact, the energy-momentum of the impinging
particles is taken to be an external light-like source --- without any
deflection --- which generates gravitational fields characterized by
Aichelburg-Sexl-like shock waves~\cite{AiSe}. However, the resulting action
predicts the Einstein deflection angle~\cite{ACV88} and corrections to
it~\cite{ACV90}, at variance with the original assumption of frozen sources.
Instead, we know from the beginning~\cite{EiIn} that in the classical limit the
gravitational equations should predict both particle motion and fields in a
self-consistent way. For that to be possible, the improved eikonal
representation is needed.

The additional important improvement of the present model is the treatment of
rescattering corrections of the produced gravitons. Such effects were argued
in~\cite{CiCoFa} to be of particular interest for the irregular solutions of the
model, perhaps explaining their relationship to collapse and their contribution
to unitarity. But, independently of such a feature, we shall argue here that
even for regular (UV-safe) solutions, rescattering corrections are needed in
order to achieve self-consistency of motion and metric at higher orders in the
eikonal expansion.

To start with, we deal with the self-consistency problem at leading level, in
which the eikonal refers to frozen undeflected sources, the metric contains two
Aichelburg-Sexl (AS) shock waves, while the action predicts the Einstein
deflection angle $\theta_E = 2R/b$, as function of the impact parameter $b$ and of
the gravitational radius $R(E) = 4GE$ ($2E=\sqrt{s}$ being the invariant mass of
the system). The problem in introducing the particle motion is that the
kinematical corrections implementing it are of the same relative order
($\theta_E^2$ or higher) as the irreducible dynamical corrections to the
Einstein deflection. How to disentangle the former terms from the latter ones?

A hint about solving the question above was provided in ref.~\cite{ACV90}, where
it was shown that --- given the particular Coulombic form of the leading eikonal
--- it is possible to take into account the motion of the Breit-frame of the
particles without affecting the $S$-matrix eigenvalue which is still provided by
the naive two-dimensional Fourier transform. Such observation is the basis for
our treatment of motion in sec.~\ref{s:ilepp}, which suggests writing a modified
metric, which is self-consistent at leading level. The main difference with the
previous one~\cite{ACV07} lies in the introduction of two-body ``shifts'' of
't~Hooft type~\cite{tH87}, summarizing the action of the leading $S$-matrix at
two-body level.

That is not enough however. The reduced-action model contains the longitudinal
fields $h^{++}$ ($h^{--}$) characterized by AS shock waves centered on the
particles with profile functions $2\pi Ra(\xt)$ ($2\pi R \bar{a}(\xt)$);
furthermore, it generates a gravitational wave also with a field $h=\nabla^2\phi$
defined by the H-diagram. The longitudinal field is here calculated in the
improved eikonal representation in sec.~\ref{s:ssw} and found to be consistent
with AS waves~\cite{AiSe} which are delayed in time and
rotated in space, as suggested in sec..~\ref{s:ilepp}. The transverse field, on the
other hand, provides corrections to the leading eikonal profiles in $a$ and
$\bar{a}$, and causes modifications of the metric inside the light-cone by
providing important rescattering corrections.

We then set up a perturbative procedure, in order to deal with the
self-consistency problem at higher orders in the eikonal expansion. At next
order we find the H-diagram deflection~\cite{ACV90} which again appears as a
feature of the action, but is not incorporated in the metric, which would
require further modifications of the energy-momentum tensor. Surprisingly, we
find that the class of multi-H diagrams considered in~\cite{ACV88} is not
sufficient in order to provide a satisfactory metric at this level but we need
to go one further step, and calculate rescattering corrections, which carry the
information due to two-body shifts for the produced gravitons. Such shifts occur
in an approximate solution to the rescattering equations under consideration
since a long time~\cite{AmCiVe}, which is here worked out to completion in
sec.~\ref{s:rcram}. The results so obtained complete the picture of the improved
eikonal representation in sec.~\ref{s:ssw} for corrections of relative order
$R^2/b^2$.

Finally, in sec.~\ref{s:isrp} we somewhat change subject, and we address the
gross features of the ultraviolet-sensitive solutions of the model, whose
importance lies in the fact that they could play a role for the recovery of
unitarity and/or the related information loss.  We then summarize the essential
results of the paper and their consequences for action and metric in
sec.~\ref{s:conc}, by discussing also some suggestions which arise from our
preliminary analysis of singular solutions.

%%%%%%%%%%%%%%%%%%%%%%%%%%%%%%%%%%%%%%%%%%%%%%%%%%%%%%%%%%%%%%%%%%%%%%%%%%%%%%%%
\section{Improved leading eikonal description and particle motion\label{s:ilepp}}
%%%%%%%%%%%%%%%%%%%%%%%%%%%%%%%%%%%%%%%%%%%%%%%%%%%%%%%%%%%%%%%%%%%%%%%%%%%%%%%%

%%%%%%%%%%%%%%%%%%%%%%%%%%%%%%%%%%%%%%%%%%%%%%%%%%%%%%%%%%%%%%%%%%%%%%%%%%%%%%%%
\subsection{Eikonal representation of the scattering amplitude\label{s:ersa}}
%%%%%%%%%%%%%%%%%%%%%%%%%%%%%%%%%%%%%%%%%%%%%%%%%%%%%%%%%%%%%%%%%%%%%%%%%%%%%%%%

\begin{figure}[t]
  \centering
  \includegraphics[width=0.7\textwidth]{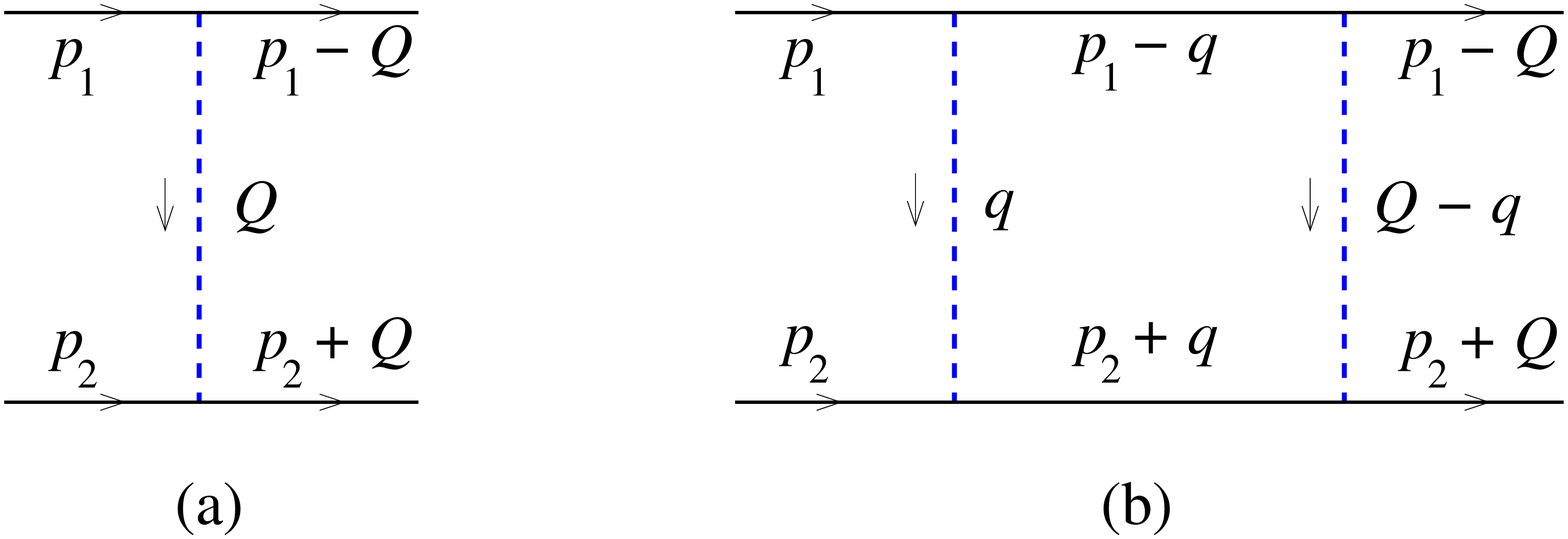}
  \Caption{One- and two-rung effective ladder diagrams determining the elastic
    $S$-matrix in the eikonal approximation. Solid lines: on-shell external
    particles; dashed lines: eikonal gravitational exchanges.}
  \label{f:ladder}
\end{figure}

ACV~\cite{ACV90} have shown that the leading contributions to the high-energy
elastic scattering amplitude come from the $s$-channel iteration of
soft-graviton exchanges, which can be represented by effective ladder diagrams
as in fig.~\ref{f:ladder}.  The purpose of the present section is to recall the
method of resumming the effective ladder contributions to all orders so as to
provide the so-called eikonal representation for the elastic $S$ matrix.  This
representation is here ``improved'' in the sense that we do not make a
separation of longitudinal and transverse variables by neglecting the leading
scattering angle. Instead, by following~\cite{ACV90}, we use the exact
3-dimensional phase-space of the on-shell particles at each gravitational
eikonal exchange and we prove that --- due to the Coulombic form of the exchange
--- the $S$-matrix is nevertheless provided by a 2-dimensional transform.

It is important to note that the exchanged gravitons (dashed lines) are associated
to a propagator $-\ui/Q^2$ and are coupled to the colliding particles/strings
(solid lines) with an interaction strength equal to $\alpha\equiv Gs/\hbar$ for
each pair of vertices.  Furthermore, the particles' lines are on-shell along the
effective ladder, as proved in ref~\cite{ACV88} by the sum over inelastic
excitations also. The generic ladder is thus built by iteration of the basic
rung
\begin{equation}\label{rung}
 \rung_1(p_1,p_2,Q) = \ui \amp_1(Q^2,s) \,
 2\pi\deltap\left((p_1-Q)^2\right)2\pi\deltap\left((p_2+Q)^2\right) \;, \quad
 \amp_1(Q^2,s)\equiv -\frac{8\pi G s^2}{Q^2}
\end{equation}

We start computing the 2-rung ladder (fig.~\ref{f:ladder}.b)
\begin{align}
 \rung_2(p_1,p_2,Q) &= \int\frac{\dif^4 q}{(2\pi)^4}\; \rung_1(p_1,p_2,q)
 \rung_1(p_1-q,p_2+q,Q-q) \nonumber \\
 &= \ui^2 \amp_2(Q^2,s) \, 2\pi\deltap\left((p_1-Q)^2\right)2\pi\deltap\left((p_2+Q)^2\right)
 \;,\label{rung2}\\
 \amp_2(Q^2,s) &\equiv \int\frac{\dif^4 q}{(2\pi)^2}\; \amp_1(q^2,s) \amp_1\left((Q-q)^2,s\right)
 \deltap\left((p_1-q)^2\right)\deltap\left((p_2+q)^2\right) \;.
 \label{amp2}
\end{align}
In the center-of-mass (CM) frame where $p_1=(E,0,0,E)$, $p_2=(E,0,0,-E)$,
$Q=(Q^0,\vQ)$, $s=4E^2$, the two mass-shell delta functions in
eq.~(\ref{amp2}) become
\begin{equation}\label{massShell}
  \deltap\left((p_1-q)^2\right)\deltap\left((p_2+q)^2\right) =
  \frac1{2s|\vq|} \delta(q^0) \delta\left(\cos\theta-\frac{|\vq|}{2E}\right)
  \;,\quad (\cos\theta=\hat{p}_1\cdot\hat{q})
\end{equation}
and, for the same reason, the two external mass-shell deltas in
eq.~(\ref{rung2}) constrain $Q^0=0$,
$\cos\theta_Q\equiv\hat{p}_1\cdot\hat{Q}=|\vQ|/2E$, $Q^2=-\vQ^2$.
In this way, the 4D integration
$\dif^4 q = \dif q^0\, |\vq|^2\; \dif|\vq|\;\dif\cos\theta\;\dif\phi$ reduces
to a 2D integral%
\footnote{This shows that $\amp_2$ is a function of $s$ and $Q^2$.}
\begin{equation}\label{a2}
  \amp_2 = \int \frac{|\vq|\dif|\vq|\dif\phi}{(2\pi)^2}\;
  \amp_1(-\vq^{\;2},s) \frac{4\pi Gs}{(\vQ-\vq)^2} \;.
\end{equation}
By taking into account the expression of $\cos\theta$ and $\cos\theta_Q$ in
terms of $|\vq|$, $|\vQ|$ and $E$ and using the identity
\begin{equation}\label{dphi}
  \int_0^{2\pi}\frac{\dif\phi}{a+b\cos\phi} = \frac{2\pi}{\sqrt{a^2-b^2}}
\end{equation}
it is easy to evaluate the azimuthal integral in eq.~(\ref{a2}) and, more
interestingly, to show that it can be rewritten as the azimuthal integral of a
pure 2D propagator:
\begin{align}\nonumber
  &\int_0^{2\pi}\frac{\dif\phi}{(\vQ-\vq)^2} = 
  \int_0^{2\pi}\frac{\dif\phi}{
    \vQ^2+\vq^{\;2}-2|\vQ||\vq|\left(\cos\theta\cos\theta_Q-\sin\theta\sin\theta_Q\cos(\phi-\phi_Q)\right)}\\
 &= \frac{2\pi}{\sqrt{(\vQ^2+\vq^{\;2})^2-4\vQ^2\vq^{\;2}}}
 = \int_0^{2\pi}\frac{\dif\phi'}{\vQ^2+\vq^{\;2}-2|\vQ||\vq|\cos\phi'}
 = \int_0^{2\pi}\frac{\dif\phi'}{(\Qt-\qt)^2} \;, \label{azimuthal}
\end{align}
where $\Qt$ and $\qt$ are fictitious 2D vectors with the same modulus as $\vQ$
and $\vq$ respectively, and $\phi'$ is the angle between them. With this
trick, $\amp_2$ can be rewritten as a 2D convolution:
\begin{equation}\label{conv2}
  \amp_2(Q^2,s) = \int\frac{\dif^2\qt}{(2\pi)^2}\;
  \amp_1(-\qt^2,s)\frac{4\pi Gs}{(\Qt-\qt)^2}
  = \frac1{2s}\left[\amp_1\otimes\amp_1\right](\Qt) \;, \quad (\Qt^2=-Q^2)\;.
\end{equation}

It is now straightforward to generalize that procedure to the $n$-rung ladder,
by deriving the recursion formula
\begin{align}
  \rung_n(p_1,p_2,Q) &= \int\frac{\dif^4 q}{(2\pi)^4}\; \rung_{n-1}(p_1,p_2,q)
 \rung_1(p_1-q,p_2+q,Q-q) \nonumber \\
 &= \ui^n \amp_n(Q^2,s) \, 2\pi\deltap\left((p_1-Q)^2\right)2\pi\deltap\left((p_2+Q)^2\right)
 \;,\label{rungn}\\
 \amp_n(Q^2,s) &= \int\frac{\dif^2\qt}{(2\pi)^2}\;
 \amp_{n-1}(-\qt^2,s) \frac{4\pi Gs}{(\Qt-\qt)^2}
 = 2s \left[\stackrel{n}{\otimes}\frac{4\pi Gs}{\Qt^2}\right] \;.
 \label{ampn}
\end{align}
It should be noted that such simplification is due to the peculiar form of the
graviton interaction from which eq.~(\ref{azimuthal}) follows.

Finally, the eikonal $S$ matrix (with overall momentum conservation factored
out) is given by the sum of the ladder diagrams with the usual combinatorial
factor $1/n!$ and with the final-state mass-shell deltas removed:
\begin{subequations}\label{Seik}
  \begin{equation}\label{Ssum}
    S_\eik(Q^2,s) = \frac1{2s}\sum_{n=0}^\infty \frac{\ui^n}{n!}\amp_n(Q^2,s)  = \sum_{n=0}^\infty
    \frac{(\ui 4\pi Gs)^n}{n!} \left[\stackrel{n}{\otimes}\frac{1}{\Qt^2}\right] \;.
  \end{equation}
  By applying a 2D Fourier transform to eq.~(\ref{Ssum}), convolutions become
  standard products and the r.h.s.\ of eq.~(\ref{Ssum}) reduces to an
  exponential series:
  \begin{equation}\label{2dft}
    \tilde{S}_\eik(\bt,s) \equiv \int\frac{\dif^2\Qt}{(2\pi)^2}\;\esp{-\ui\Qt\cdot\bt}
    S_\eik(-\Qt^2,s) = \exp\{\ui\Aeik(\bt,s)\} \;, \qquad \Aeik\equiv \tilde{\amp}_1
  \end{equation}
\end{subequations}
where we interpret the 2D Fourier transform of the one-graviton amplitude
$\tilde{\amp}_1$ as {\em eikonal action} $\Aeik$.
Actually, the definition of $\Aeik$ (and of all convolutions involved
in this calculation) requires the introduction of an infrared cutoff
$Q_0\sim 1/L$ in order to regularize the ``Coulomb'' divergence typical of
long-range interactions $\sim1/Q^2$:
\begin{equation}\label{amp1b}
 \Aeik(\bt,s) \equiv 2\delta_0(b,s) = \int\frac{\dif^2\Qt}{(2\pi)^2}\;
 \esp{-\ui\Qt\cdot\bt} \frac{4\pi Gs}{\Qt^2} \Theta(\Qt^2-Q_0^2)
 = 2 Gs\ln\frac{L}{b} + \ord{\frac{b}{L}}^2
\end{equation}
where $b\equiv|\bt|$ and $L\equiv 2\esp{-\gamma_E}/Q_0$.

A couple of remarks are in order:
\begin{itemize}
\item The ``position'' variable $\bt$, being conjugate to $\Qt$, is usually
  interpreted as impact parameter of the collision, and $\delta_0(b,s)$ as the
  associated phase shift. However, $\Qt$ is a fictitious 2D vector whose polar
  angle is actually undefined, therefore $\bt$ cannot be thought of as a purely
  transverse (to $\vec{p}_1$) vector.
\item The dependence of $\tilde{S}$ on $L$ amounts to a $b$-independent phase,
  and therefore it is unimportant in the determination of the scattering
  angle. However such a phase depends on the total energy $\sqrt{s}$. This may
  cause a cutoff-dependent time evolution  of the scattered particles, hopefully
  without physical consequences.
\end{itemize}
Both issues will be discussed in detail in the next section.

%%%%%%%%%%%%%%%%%%%%%%%%%%%%%%%%%%%%%%%%%%%%%%%%%%%%%%%%%%%%%%%%%%%%%%%%%%%%%%%%
\subsection{Wave packet motion induced by the $\bs{S}$-matrix\label{s:wpm}}
%%%%%%%%%%%%%%%%%%%%%%%%%%%%%%%%%%%%%%%%%%%%%%%%%%%%%%%%%%%%%%%%%%%%%%%%%%%%%%%%

The purpose of this section is to derive the motion of two quantum particles
subject to a scattering amplitude given by the leading eikonal result
of~\cite{ACV90} and recalled in sec.~\ref{s:ersa}. Suppose we prepare a state of
two well separated light-like free particles in the past. In a
first-quantization description%
\footnote{ Since we are neglecting particle production, the first-quantization
  framework is suited for our purposes.}
the wave function is given by
\begin{equation}\label{d:psiIn}
  \psi_\in(t,\vx_1,\vx_2) = \int\dif^3 p_1 \dif^3 p_2\;\tilde\psi(\vp_1,\vp_2)
  \esp{-\ui E_1 t+\ui\vp_1\cdot\vx_1}\esp{-\ui E_2 t+\ui\vp_2\cdot\vx_2} \;,
  \qquad(E_i = |\vp_i|) \;.
\end{equation}
In the far future the wave function is expressed in terms of the $S$-matrix as
follows:
\begin{align}\label{d:psiOut}
  \psi_\out(t,\vx_1,\vx_2) &= \int\dif^3 p_1 \dif^3 p_2\dif^3 k_1 \dif^3 k_2\;
  \tilde\psi(\vp_1,\vp_2) \langle \vk_1,\vk_2|S|\vp_1,\vp_2\rangle
  \esp{-\ui k_1 x_1}\esp{-\ui k_2 x_2} \\ \nonumber
  k_j &= (E'_j,\vk_j) \;,\qquad E'_j = |\vk_j| \;.
\end{align}

It is convenient to perform the calculation in the CM frame. As usual, we define
the CM coordinate $\vec{X}$ and relative coordinate $\vx$
\begin{subequations}\label{CMcoordinates}
\begin{align}
 \vec{X} &\equiv \frac{\vx_1+\vx_2}{2} && \text{conjugated to total momentum} &&
 \vec{P}\equiv \vp_1+\vp_2 \\
 \vx &\equiv \vx_1-\vx_2 && \text{conjugated to relative momentum} && \vp \equiv
 \frac{\vp_1-\vp_2}{2}
\end{align}
\end{subequations}
For simplicity, we limit ourselves to states which have zero total momentum, by
setting
\begin{subequations}\label{wavefunc}
  \begin{equation}\label{cmCond}
    \tilde\psi(\vp_1,\vp_2) = \delta^3(\vp_1+\vp_2) \tilde\psi(\vp) \;.
  \end{equation}
With this restriction the CM coordinate is completely delocalized, but we can
build a state representing a localized wave packet in the relative coordinate:
  \begin{equation}
    \tilde\psi(\vp) = \tilde{N}\exp\left\{-\frac12\left(\frac{\vp-\vp_0}{\sigma}\right)^2
      -\ui\vp\cdot\vx_0 + \ui 2E_p t_0 \right\} \;,
  \end{equation}
\end{subequations}
where $\vp=\vp_1=-\vp_2$ is the relative momentum, $\vp_0=(0,0,E)$ is its mean
value and $\sigma$ a measure of its broadening, $\tilde{N}$ is a normalization
factor, $2E_p=2|\vp|$ is the total CM energy, while $\vx_0$ is the relative
position at time $t_0$.  In fact, by inserting eqs.~(\ref{wavefunc}) into
eq.~(\ref{d:psiIn}) we are left with an integral in $\vp$ whose phase $\Phi_\in
= -2E_p (t-t_0)+\vp\cdot(\vx-\vx_0)$ is stationary for
\begin{equation}\label{xIn}
  \vx_\in-\vx_0 = 2 \nabla_{\vp}E_p \, (t-t_0) \simeq 2\frac{\vp_0}{E}(t-t_0)
\end{equation}
since the shape of $\tilde\psi$ forces $\vp\simeq \vp_0$. Going back to
particles' coordinates, we confirm that in the past each particle moved at the
speed of light in the direction of $\vp_0$:
\begin{equation}\label{x1In}
 \vx_1(t) = -\vx_2(t) = \frac{\vx}{2} = \frac12 \vx_0 + \hat{p}_0(t-t_0) \;, \qquad \hat{p}_0
 \equiv \frac{\vp_0}{E} = (0,0,1)\;.
\end{equation}

On the other hand, by substituting eqs.~(\ref{wavefunc}) in the outgoing
wave~(\ref{d:psiOut}) together with the energy-momentum conserving eikonal
$S$-matrix~(\ref{Seik})
\begin{align}
  \langle \vk_1,\vk_2|S|\vp_1,\vp_2\rangle &= \delta^4(k_1+k_2-p_1-p_2)\,2 s 
  \int \dif^2 \bt \; \esp{i\Qt\cdot\bt+i\Aeik(\bt,s)} \\
  Q&\equiv p_1-k_1 =  p-k\;, \qquad \Qt^2 \equiv -Q^2 \;, \qquad s = \big(2E_p \big)^2
\end{align}
we find
\begin{align}\label{psiOut}
  \psi_\out &= \half\tilde{N}\int\dif^3 p\, \dif^3 k\, \dif^2 \bt \; \delta(E_k-E_p)
  \exp\left\{-\frac12\left(\frac{\vp-\vp_0}{\sigma}\right)^2\right\}\exp(\ui\Phi_\out) \\
  \Phi_\out &= -2E_p(t-t_0) + \vk\cdot\vx - \vp\cdot\vx_0 + \Qt\cdot\bt +
  2Gs\log\frac{L}{b} \;,
\end{align}
where $\Qt$ and $\bt$ are auxiliary 2D vectors, the former being determined by the
condition $\Qt^2=-Q^2$.  In this case, the stationarity
of $\Phi_\out$ with respect to independent variations of $\vp$, $\vk$ and $\bt$
yields the asymptotic motion in the future, as follows.  In the plane of
scattering $\langle\vp,\vk\rangle$ we introduce the polar angles $\theta_i$ of
momenta and coordinates, and write
\begin{equation}\label{polar}
  \vk\cdot\vx   = E_p \,r   \cos\theta_{kx}   \;, \quad
  \vp\cdot\vx_0 = E_p r_0 \cos\theta_{px_0} \;, \quad
  \Qt\cdot\bt = |\vQ|\,b\cos\phi' = 2E_p\left|\sin\frac{\theta_{kp}}{2}\right| b \cos\phi'
\end{equation}
where we used the notation $\theta_{ij}\equiv\theta_i-\theta_j$, $b\equiv|\bt|$,
$r\equiv|\vx|$ and the relations $E_k=E_p$, $|\vQ|=2E_p|\sin(\theta_{kp}/2)|$.

\begin{figure}[t]
  \centering
  \includegraphics[width=0.6\textwidth]{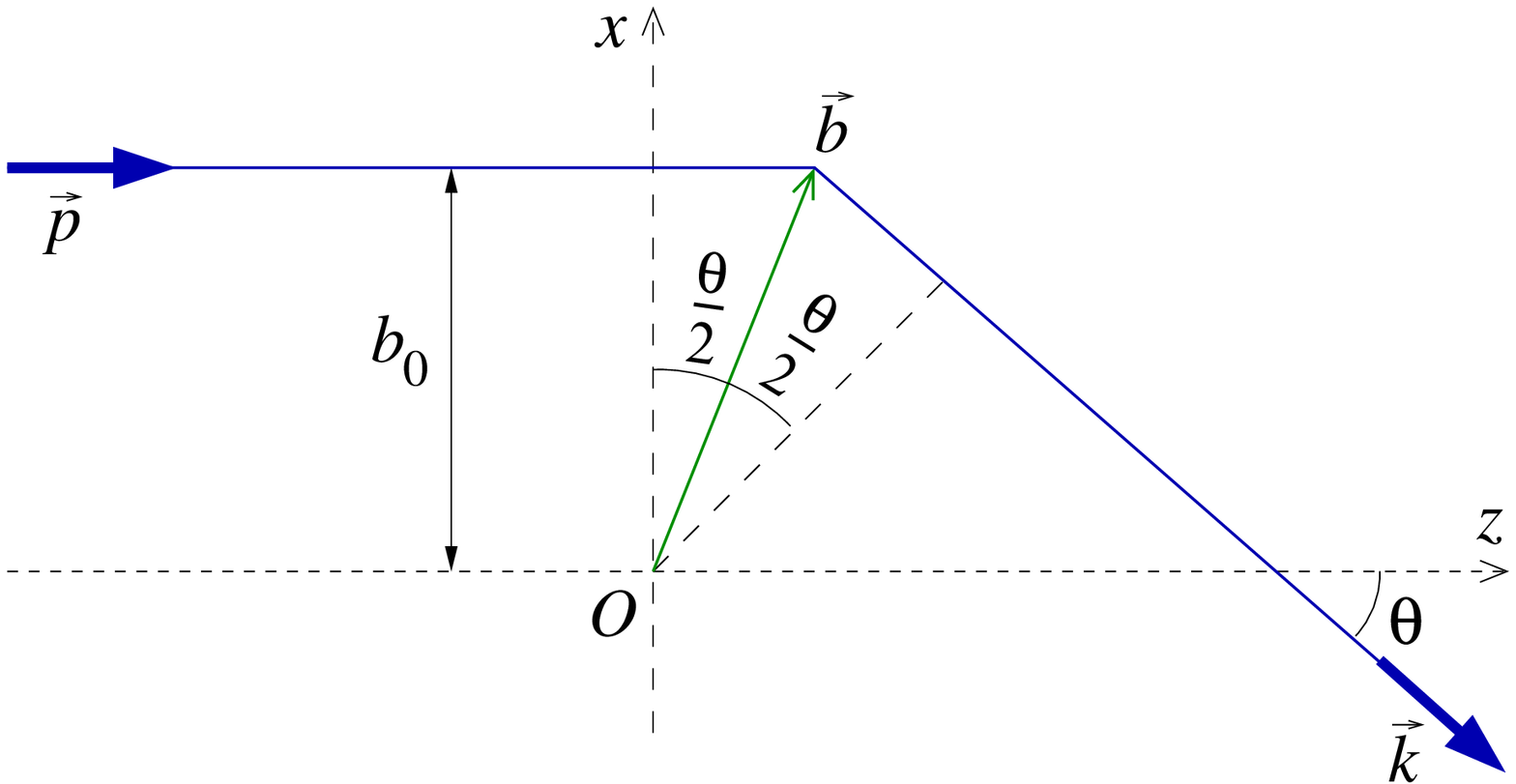}
  \Caption{Kinematics of the wave packets' relative coordinates.}
  \label{f:deflection}
\end{figure}

We can now determine the five unknowns $r$, $\theta_k$, $\theta_p$ and $\bt$ by
finding the stationarity point of the phase with respect to variations of the five
integration variables $E_p$, $\theta_k$, $\theta_p$ and $\bt$, while $t_0$ and
$\vx_0$ are known parameters. We have
\begin{align}
  0 &= \frac{\partial\Phi_\out}{\partial\bt} = \Qt + \frac{\partial\Aeik}{\partial\bt}
   = \Qt - \hat{b} \frac{2Gs}{b} \label{eqb} \\
  0 &= \frac{\partial\Phi_\out}{\partial\theta_p} = E_p r_0 \sin\theta_{px_0}
   -\sgn(\theta_{kp}) \; E_p b \cos\frac{\theta_{kp}}{2} \cos\phi' \label{eqp} \\
  0 &= \frac{\partial\Phi_\out}{\partial\theta_k} = -E_p r \sin\theta_{kx}
   +\sgn(\theta_{kp}) \; E_p b \cos\frac{\theta_{kp}}{2} \cos\phi' \label{eqk} \\
  0 &= \frac{\partial\Phi_\out}{\partial E_p} = -2(t-t_0)+r\cos\theta_{kx}
   -r_0\cos\theta_{px_0}+2\left|\sin\frac{\theta_{kp}}{2}\right|b\cos\phi'
   +8E_p\frac{\partial\Aeik}{\partial s} \;, \label{eqE}
\end{align}
where $\sgn(x)$ denotes the sign of $x$.
The first equation causes $\bt$ to be aligned with $\vQ$ ($\phi'=0$) and to have modulus
\begin{equation}\label{solb}
  b = \frac{2Gs}{|\vQ|} = \frac{R}{\left|\sin\frac{\theta_{kp}}{2}\right|} \;.
\end{equation}
The second equation yields a further relation involving the physical impact
parameter $b_0$:
\begin{equation}\label{b0}
  b_0 \equiv r_0\sin\theta_{x_0\,p} = -\sgn(\theta_{kp}) \;b \cos\frac{\theta_{kp}}{2}
  = b\cos\frac{\theta}{2}
\end{equation}
allowing us to determine the (attractive) scattering angle
$\theta\equiv-\theta_{kp}>0$
\begin{equation}\label{angle}
  \tan\frac{\theta}{2} = \frac{R}{b_0} \;.
\end{equation}
The third equation identifies the outgoing trajectory to be the straight line
\begin{equation}\label{outTraj}
  r\sin\theta_{xk} = b\cos\frac{\theta}{2} = b_0
\end{equation}
with the same impact parameter of the incoming one, thus implying angular
momentum conservation. Note that this trajectory intersects the incoming one at
the point $\vb=b\big(\cos\theta/2,0,\sin\theta/2)\big)$, as is apparent from
fig.~\ref{f:deflection}.
The fourth equation provides the motion of the outgoing relative coordinate
\begin{equation}\label{motion}
  \frac12 r\cos\theta_{xk} = (t-t_0) + \frac12 r_0\cos\theta_{x_0\,p}-b\sin\frac{\theta}{2}
  - 2R\ln\frac{L}{b}
\end{equation}
meaning that each particle keeps on moving at the speed of light.

A very peculiar behaviour of our result is that the motion of the outgoing
particles suffers a time delay given by the energy-derivative of the action
\begin{equation}\label{delay}
  t_\delay = \frac12 \frac{\partial\Aeik}{\partial E_p} = 2R\ln\frac{L}{b}
\end{equation}
with respect to a motion with the same deflection but at constant speed. This is
easily seen by projecting the incoming trajectory~(\ref{xIn}) on $\vp$
\begin{equation}\label{inMotion}
  \frac12 r_\in\cos\theta_{xp} = \frac12 r_0\cos\theta_{x_0\,p} + (t-t_0) \;,
  \qquad r_\in\sin\theta_{x\,p} = b_0
\end{equation}
and by choosing the origin of time $t_0=0$ when $\vx_\in$ is (or would be) at the
intersection point $\vb$ of the trajectories~(\ref{outTraj},\ref{inMotion}), i.e.\
$r_0\cos\theta_{x_0\,p}=b\sin(\theta/2)$, so that
eqs.~(\ref{inMotion},\ref{motion}) become
\begin{align}
  \frac12 r_\in \cos\theta_{xp}  &= t+\frac{b}{2} \sin\frac{\theta}{2} \\
  \frac12 r_\out \cos\theta_{xk} &= t-t_\delay - \frac{b}{2} \sin\frac{\theta}{2}
\end{align}
showing that $\vx_\out=\vb$ at $t=t_\delay$. Let us stress that this time delay
is peculiar of gravity, since the energy dependence of the amplitude stems from
the gravitational coupling $\alpha=Gs$. In other theories, like electrodynamics,
the coupling is independent%
\footnote{Or very weakly dependent after renormalization.}
of the particles' energies, and one obtains analogous equations with
$t_\delay = 0$.

%%%%%%%%%%%%%%%%%%%%%%%%%%%%%%%%%%%%%%%%%%%%%%%%%%%%%%%%%%%%%%%%%%%%%%%%%%%%%%%%
\subsection{Scattering description by trajectory shifts\label{s:sdts}}
%%%%%%%%%%%%%%%%%%%%%%%%%%%%%%%%%%%%%%%%%%%%%%%%%%%%%%%%%%%%%%%%%%%%%%%%%%%%%%%%

We have seen that the leading eikonal $S$-matrix provides a definite scattering
angle and wave-packet motion. It also provides a gravitational metric,
associated to the gravitaional fields $h_{\mu\nu}=g_{\mu\nu}-\eta_{\mu\nu}$,
whose longitudinal terms are expressed in terms of the eikonal amplitude $a(\xt)$ as
follows~\cite{ACV07}
\begin{subequations}\label{hFields}
\begin{align}
  \fourth h^{++} = h_{--} &= 2\pi R a(\xt) \delta(x^-) \;,&
   x^+ &\equiv t+z \label{h--}\\
  \fourth h^{--} = h_{++} &= 2\pi R \bar{a}(\xt) \delta(x^+) \;,&
   x^- &\equiv t-z \;, \label{h++}
\end{align}
\end{subequations}
where
\begin{equation}\label{a0}
  a(\xt) = \frac1{2\pi}\ln\frac{L^2}{\xt^2} \;, \qquad \bar{a}(\xt) = a(\bt-\xt)
\end{equation}
and we actually deal with the situation before collision ($t<0$).

The collision process, according to the reduced action model to be described in
sec.~\ref{s:rcram}, introduces further (transverse) components of the metric
field, and modifies the expressions (\ref{hFields}) by corrections of relative
order $R^2/b^2$ and higher. Nevertheless, at leading level, starting from the
(improved) eikonal representation only, it should be possible to describe the
scattering on the basis of an associated metric. Here we suggest a simple way to
do it, which is based on 't~Hooft's understanding~\cite{tH87} of the $S$-matrix
as a coordinate shift.

The starting point is to write the leading order metric before collision,
provided by the expressions~(\ref{hFields}) as follows:
\begin{equation}\label{Lmetric}
  \dif s^2 = \dif \xt^2 - \dif x^- \dif x^+ + 2\pi R \left[
    a(\xt) \delta(x^-) (\dif x^-)^2 + \bar{a}(\xt) \delta(x^+) (\dif x^+)^2\right] \;.
\end{equation}
We see that it consists of two Aichelburg-Sexl (AS) shock waves~\cite{AiSe}
travelling against each other. If only one is present --- the one at $x^+ =0$, say
---, a test particle impinging on it will acquire a shift in the $x^-$ direction:
\begin{equation}\label{vshift}
  \Delta x^- = 2\pi R\, \bar{a}(\xt)\;, \qquad \Delta x^+ = 0\;, \qquad\imp\qquad
  \Delta t = -\Delta z = \pi R\, \bar{a}(\xt) \;.
\end{equation}
As a consequence, after the shift, the test particle will be deflected. In
particular, if the particle was moving at fixed $x^-$ and $\xt$, the scattering
angle $\theta(\xt)$ is given by $\tan(\theta/2) = \pi R |\bs{\nabla} \bar{a}(\xt)|$,
see fig.~\ref{f:deflection}.

The scattering angle formula is derived by either calculating the action
$2\pi G s \bar{a}(\xt)$ of eq.~(\ref{amp1b}) from the shift~(\ref{vshift}) as in
ref.~\cite{ACV07} and then using eq.~(\ref{angle}), or by computing the
geodesics by standard methods.

\begin{figure}[ht!]
  \centering
  \includegraphics[width=0.6\textwidth]{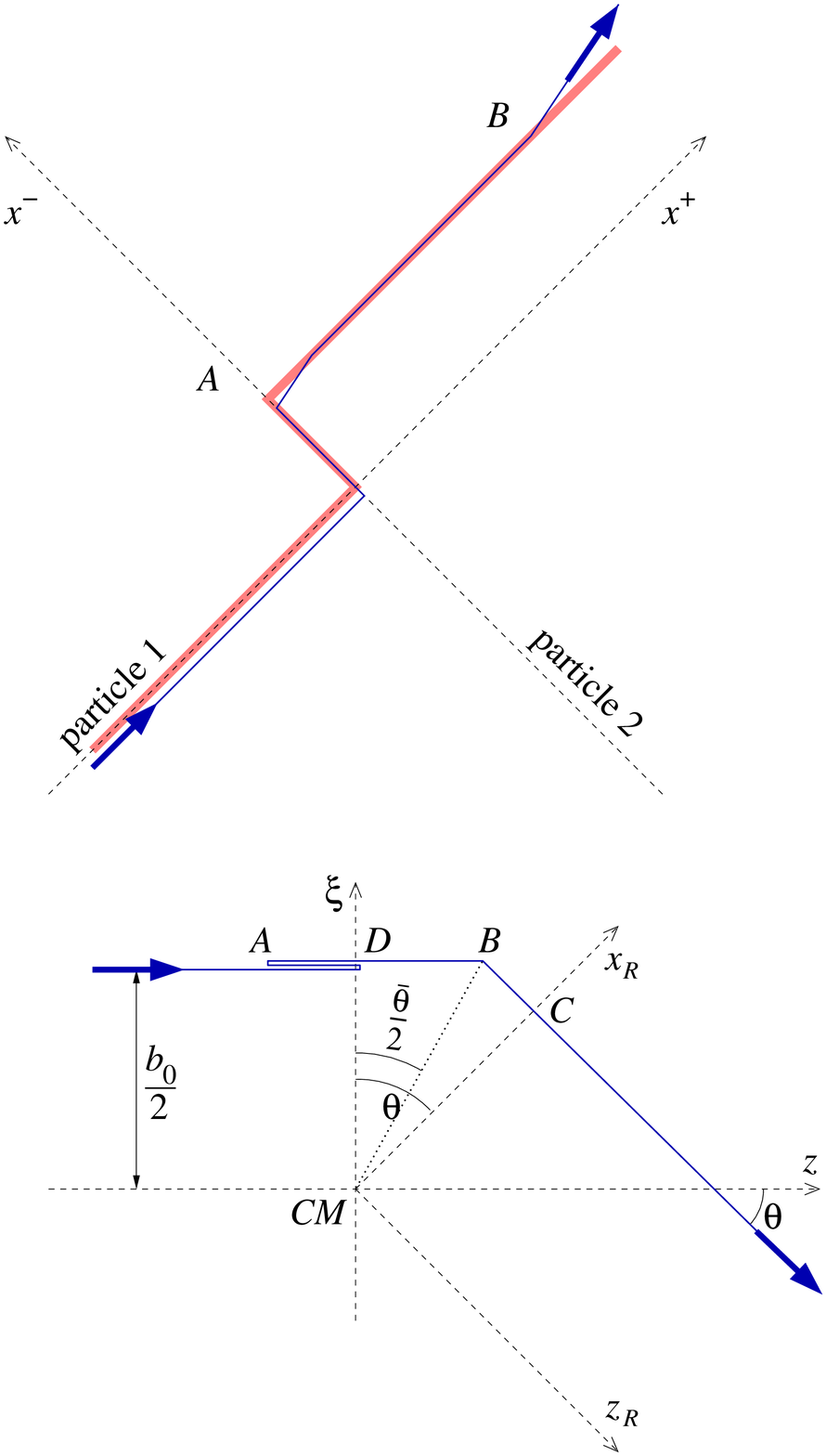}
  \Caption{Space-time diagram of the double-shift picture of a test particle
    (solid blue) by the shock-wave of particle 2 (not shown) and then by the
    shifted shock wave of particle 1 (thick pink line). In the upper part, the
    motion in the longitudinal plane is shown. In the lower part, the coordinate
  $\xi$ denote the transverse direction starting from the center of mass (CM) of
  the two colliding particles. At the point $A$ ($B$) the test particle leaves the
  first (second) wave front.}
  \label{f:doppioShift}
\end{figure}

The next question is: what happens when both AS waves are present? In order to
understand that, consider a test particle at $x^- =-\e < 0$, travelling along
with particle 1 ($x^- =0$) shortly before it. Then, at $x^+ =0$ ($t=-z=-\e/2$)
the test particle will collide with particle 2 and will acquire the
shift~(\ref{vshift}), while the location of particle 1 will be forced to do the
same with $\Delta x^-_1 = 2\pi R \bar{a}(0)$, evaluated at the ``on shell''
point $\xt\equiv \bt_0/2+\xi = 0$ ($\xi = -\bt_0/2$), see
fig.~\ref{f:doppioShift}. Next, the test particle trajectory is deflected on a
trajectory $z\simeq t \cos\theta(x)$, $x\simeq -t \sin\theta(x)$, therefore, due
to the $\cos\theta$ factor, it will be shortly hit by particle 1, travelling at
$t-z=\text{const}$, and thus will acquire the shift
\begin{equation}\label{ushift}
  \Delta x^+ = 2\pi R a(x) \;, \qquad \Delta x^- = 0 \qquad\imp\qquad
  \Delta t = \Delta z = \pi R a(x) \;.
\end{equation}
To sum up, the test particle acquires two shifts
\begin{equation}\label{uvshift}
  \Delta x^- = 2\pi R \bar{a}(x) \quad\text{at}\quad x^+ =0 \;, \qquad
  \Delta x^+ = 2\pi R      a (x) \quad\text{at}\quad x^- =2\pi R \bar{a}(x)
\end{equation}
separated by a short time interval $\ord{\e}$, which vanishes in the
$\e\to0$ limit. As a consequence, it is natural to assume that particle 1
acquires two shifts also:
\begin{equation}\label{uvshift2}
  \Delta x^-_1 = 2\pi R \bar{a}(0) \quad\text{at}\quad x^+ =0 \;, \qquad
  \Delta x^+_1 = 2\pi R \tilde{a}  \quad\text{at}\quad x^- =2\pi R \bar{a}(0) \;,
\end{equation}
where $\tilde{a}\equiv a(0)$ is actually a possibly singular parameter
describing the scattering of particle 1 on itself, which is here taken to be a
$b_0$-independent constant, $\tilde{a} > \bar{a}(0)$.

Taking into account both shifts, particle 1 goes from $z=0^-$ to
$\Delta z = \pi R(\tilde{a}-\bar{a})$ at time
$\Delta t = \pi R(\tilde{a}+\bar{a}) = 2\pi R \bar{a} + \Delta z \equiv t_D + \Delta z$
and is then deflected by an angle $\theta(\xt=0)$ on the trajectory%
\footnote{The subscript $R$ means ``rotated''; see also eqs.~(\ref{rotation}).}
\begin{subequations}\label{cooR}
\begin{align}
  z_R(t) &= \Delta z \cos\theta - \xi \sin\theta + (t-\Delta t) \label{zR}\\
  x_R(t) &= \xi \cos\theta + \Delta z \sin\theta = \text{const} \label{xR} \;,
\end{align}
\end{subequations}
where we recall that $x = b_0/2+\xi$. The total collision time --- from $z=0^-$
to $z_R(t_C)=0$ --- is therefore
\begin{equation}\label{collisionTime}
  t_C = \pi R (\tilde{a}+\bar{a}) - \Delta z \cos\theta + \xi \sin\theta
      = 2\cos^2(\halftheta)\,\bar{a}(x) + 2\sin^2(\halftheta) \,\tilde{a}
      + 2\sin(\halftheta)\cos(\halftheta)\,\xi \;.
\end{equation}
From the stationarity condition of the collision time with respect to $\xi$, we
obtain the scattering angle
\begin{equation}\label{scatAng}
  \tan\frac{\theta(x)}{2} = \pi R\left|\frac{\dif \bar{a}(x)}{\dif x}\right|_{x=0}
  = \pi R\left|\frac{\dif a(b_0-x)}{\dif x}\right|_{x=0} = \frac{R}{b_0}\;,
\end{equation}
where $x=b_0/2+\xi = 0$ is ``on-shell''. Since $b_0 = b\cos(\theta/2)$,
eq.~(\ref{scatAng}) is consistent with the action determination, at leading
level:
\begin{equation}\label{thetaL}
  \sin\frac{\theta}{2} = \pi R \left|\frac{\dif a(b)}{\dif b}\right|
  = \frac{R}{b} \;, \qquad\imp\qquad \tan\frac{\theta}{2} = \frac{R}{b_0} \;.
\end{equation}

Besides providing the scattering angle, the double-shift picture has a strict
analogy with the wave packet motion in sec.~\ref{s:wpm}. The former can be
summarized in 3 steps:
\begin{itemize}
\item[a)] The ``come-back'' motion from $z=0^-$ back to $z=0^+$, taking the time
  $t_D = 2\pi R\, \bar{a}(0) = 2\pi R\, a(b)$ which is just the retardation time
  with respect to travel at the speed of light;
\item[b)] the motion along the residual shift $\Delta z$, taking time
  \begin{equation}\label{timeB}
  \Delta z = \pi R\left[\tilde{a} - \bar{a}\left(\textstyle{\frac{b}{2}}+\xi\right)
   \right] = \xi\tan\frac{\bar\theta}{2}\;,
\end{equation}
where we have defined the angular shift $\bar\theta$ as in fig.~\ref{f:doppioShift};
\item[c)] The motion along the deflected trajectory, taking time
\begin{equation}\label{timeC}
  \xi\sin\theta -\Delta z \cos\theta =
  \xi\sin\left(\theta-\frac{\bar\theta}{2}\right) \Big/
  \cos\left(\frac{\bar\theta}{2}\right) \,.
\end{equation}
\end{itemize}
By comparison, in the wave packet motion the delay time is just the same, while
steps (b) (travel to point $\vx_0=\vb$ joining incoming and outgoing
trajectories) and (c) (travel to $z_R=0$) both take time $\xi\tan(\theta/2)$.

Although different, the expressions~(\ref{timeB}) and (\ref{timeC}) correspond
to the total time
\begin{equation}\label{totTime}
 2\xi\tan\frac{\theta}{2}
 \left[1+\sin\left(\frac{\theta}{2}\right)\sin\left(\frac{\theta-\bar\theta}{2}\right)
   \Big/ \cos\left(\frac{\bar\theta}{2} \right) \right]
\end{equation}
whose difference with the former is of relative order $R^2/b^2$, because both
$\theta$ and $\bar\theta$ are of order $R/b$. Therefore the two pictures coincide
at leading order.

Similarly, the outgoing impact parameter is
\begin{equation}\label{outIF}
  \xi\cos\theta+\sin\theta\,\Delta z = \xi
  \left[1-2\sin\left(\frac{\theta}{2}\right)\sin\left(\frac{\theta-\bar\theta}{2}\right)
    \Big/ \cos\left(\frac{\bar\theta}{2}\right) \right]
\end{equation}
which differs from the incoming one by relative order $R^2/b^2$
also. Furthermore both differences vanish in the limit $\theta = \bar\theta$,
which is obtained by setting
\begin{equation}\label{linearModel}
  a\left(\textstyle{\frac{b_0}{2}}+\xi\right) = \tilde{a} + a'\,\xi \;, \qquad
  \text{with $\tilde{a}$ and $a'$ constants.}
\end{equation}
This model, that we could call the ``linear model'', could be described by the
scattering of extended sources, in which the energy of each source inside a
disk of radius $r$ increases like $r$.

We conclude that the double-shift picture provides us with some understanding of
the evolution of the metric at collision and after, which is satisfactory at
leading level, even if a full account of the collision process is lacking.
The basic features argued so far are:
\begin{itemize}
\item The existence of a delay time $t_D=2\pi R\, a(b)$ which causes a shift of
  either shock wave around $z=0$ in the form
  \begin{equation}\label{h++shift}
    h_{++} = 2\pi R\, \bar{a}(x) \, \delta\big(x^+ -2\pi R\, a(b)\Theta(t-t_D)\Theta(t_C-t)\big)
  \end{equation}
around particle 1, with a similar one for particle 2;
\item The deflection of the particle trajectories, causing a rotation by the
  scattering angle $\theta$ for $t>t_B$ and reaching $z_R=0$ at the collision
  time $t=t_C$
  \begin{align}
    h_{++} &= 2\pi R \bar{a}(\xt_R)\delta\left(x^+_R-2\pi R\, a(b)\Theta(t-t_C)\right) \nonumber\\
    h_{--} &= 2\pi R a(\xt_R) \delta\left(x^-_R-2\pi R\,\bar{a}(b)\Theta(t-t_C)\right) \;,
    \label{hR}
  \end{align}
  where the rotated coordinates are
  \begin{align}
    x^{\pm}_R &= t \pm z \cos\theta \nonumber \\
    \xt^1_R &= \xt^1 \cos\theta - \xt^2 \sin\theta \nonumber \\
    \xt^2_R &= \xt^1 \sin\theta + \xt^2 \cos\theta \;. \label{rotation}
  \end{align}
\end{itemize}
Therefore, the particle motion occurs explicitly in the expression of the metric,
that we could call at this point, a self-consistent leading metric.

%%%%%%%%%%%%%%%%%%%%%%%%%%%%%%%%%%%%%%%%%%%%%%%%%%%%%%%%%%%%%%%%%%%%%%%%%%%%%%%%
\section{Shifted shock-wave fields from the improved\\ eikonal model\label{s:ssw}}
%%%%%%%%%%%%%%%%%%%%%%%%%%%%%%%%%%%%%%%%%%%%%%%%%%%%%%%%%%%%%%%%%%%%%%%%%%%%%%%%

\begin{figure}[t!]
  \centering
  \includegraphics[width=0.9\textwidth]{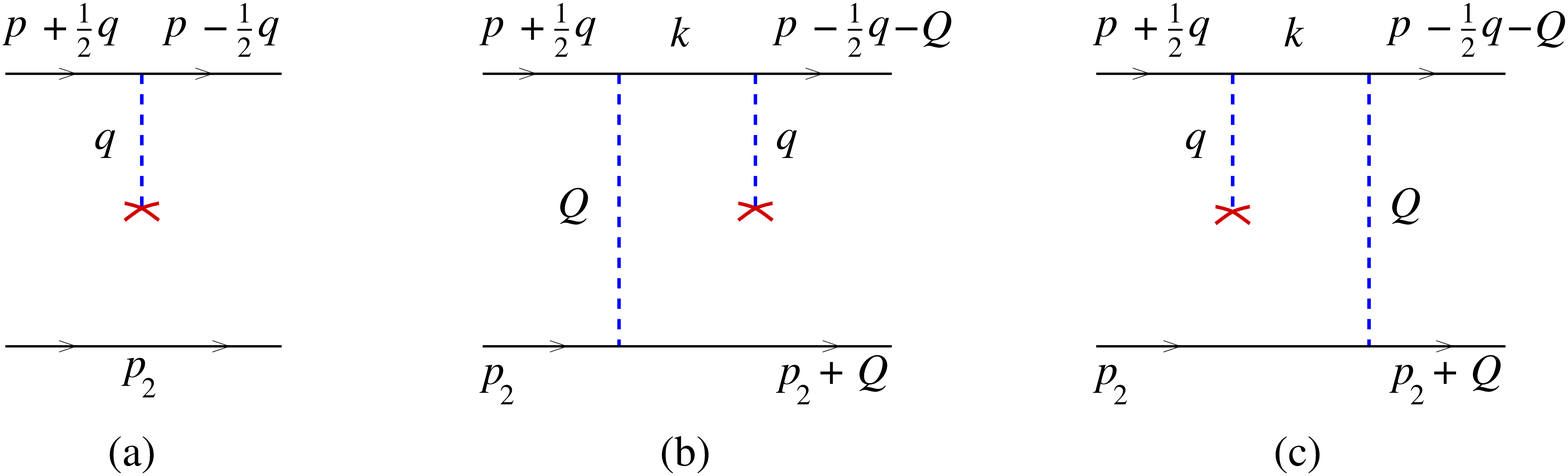}
  \Caption{Simplest eikonal diagrams with one insertion of the $h_{--}$ field
    (red cross). The parametrization of momenta is choosen so as to simplify the
    the calculations.}
  \label{f:simpleInsertion}
\end{figure}

We have seen before how the improved eikonal model is able to describe particle
motion at leading level~(sec.~\ref{s:wpm}), and how this feature can be
incorporated by trajectory shifts in the corresponding shock-waves
(sec.~\ref{s:sdts}). We want now to show how the shock-wave fields can be
explicitly derived from the sum of diagrams describing eikonal scattering, as
defined in sec.~\ref{s:ersa}. We confirm in this way the delay shifts
introduced previously, leading eventually to the scattered energy-momentum.

In order to calculate the fields $h_{++}$ and $h_{--}$ we should consider the
eikonal diagrams (fig.~\ref{f:ladder}) in presence of a linearly coupled
external source ($T^{++}$, $T^{--}$) which generates the fields by a functional
derivative of the corresponding semiclassical $S$-matrix or, in other words, of
the corresponding action. We can write
\begin{equation}\label{derFunz}
  h_{--}(x) = S^{-1} \left.\frac{\delta S}{\delta T^{--}(x)}
  \right|_{T^{--}=0}
\end{equation}
and compute the result by inserting the $T^{--}$ source on the eikonal diagrams,
in all possible ways.

Since the external current diagrams generate insertions of momentum $q$ and thus
generally do not conserve energy-momentum, we should restore time-ordered
integrations in order to evaluate them properly.  Since $h_{--}$ is coupled to
$p^{+}$ only, the $p^{-}$ conservation is not affected, and it is sufficient to
consider $x^+$-ordered diagrams as depicted in fig.~\ref{f:simpleInsertion}. Let
us first consider the small scattering angle kinematics $\theta \ll 1$, in which
the longitudinal dynamics is separated from the transverse one.

In order to understand the issue, let us start from the simplest diagrams in
fig.~\ref{f:simpleInsertion}. We notice that, while the $Q$-exchange in the
eikonal line has negligible $Q^{\pm}=\ord{Q^2/\sqrt{s}}$ because of two
mass-shell conditions $(p_1-Q)^2 = 0 = (p_2+Q)^2$, the $q$-exchange has only
one, $(k-q)^2=0$, so that some $q^+ \gtrsim |\qt|$ is still allowed, provided
$q^+ \ll \sqrt{s}$. Therefore $q^+ q^- = \ord{\qt^2 q^+/\sqrt{s}}$ is still
negligible, but the $q^+$ leakage should be considered in the $p^+$ evolution.

In the example of fig.~\ref{f:simpleInsertion} we choose to couple the $T^{--}$
current to initial and final particles of type 1 symmetrically,%
\footnote{Furthermore, we shall weight the initial and final states
  in~(\ref{derFunz}) with a wave function having energy centered around
  $p^+=\sqrt{s}=p_2^-$, which amounts to setting $p_1^+=\sqrt{s}+\half q^+$ and
$p_1'^+=\sqrt{s}-\half q^+$.}
so that a $q^+$-independent factor of $\kappa^2 p^+/(2\qt^2)$ is factored out.
In this way, the lowest-order diagram in fig.~\ref{f:simpleInsertion}.a yields
the Born-level field
\begin{equation}\label{h0++}
  h_{--}^{(0)}(x^+,x^-,\xt) = 4\pi G\sqrt{s} \int \frac{\dif^2\qt}{(2\pi)^2} \;
 \frac{\esp{\ui \qt\cdot\xt}}{\qt^2} \int \frac{\dif q^+}{4\pi}\;
 \esp{-\ui\frac{q^+}{2} x^-} = 2\pi R a_0(\xt) \delta(x^-) \;,
\end{equation}
where $a_0(\xt) = \frac{1}{2\pi}\log\frac{L^2}{\xt^2}$ is the leading profile
function of the shock-wave.

On the other hand, the $Q$-exchange yields the leading eikonal $\ui a_0(\bt)$
coupled to the charge
$2\pi G \bar{p}^- (p^+ +\frac12 q^+) = 2\pi G(s+\sqrt{s} \frac12 q^+)$, where
$\frac12 q^+ = \ui\partial/\partial x^-$ according to eq.~(\ref{h0++}). Therefore,
diagram~\ref{f:simpleInsertion}.b predicts a (subleading) contribution to
$h_{--}$
\begin{subequations}\label{h1bc}
\begin{equation}\label{h1b++}
  h_{--}^{(1b)} = 2\pi R a_0(\xt)\left[ -\pi R\delta'(x^-) a_0(\bt)\Theta(x^+)
  \right]
\end{equation}
while diagram~\ref{f:simpleInsertion}.c yields
\begin{equation}\label{h1c++}
  h_{--}^{(1c)} = 2\pi R a_0(\xt)\left[ \pi R\delta'(x^-) a_0(\bt)\Theta(-x^+)
  \right]
\end{equation}
\end{subequations}
because of the opposite sign in $q^+$ and $x^+$.

We shall interpret the results~(\ref{h1bc}) as the first order expansion in
$R/x^-$ of a shifted field
\begin{equation}\label{h++sh}
  h_{--}(x^+,x^-,\xt) = 2\pi R a_0(\xt)\delta\left(x^- - 2\pi R \half\sgn(x^+) a_0(\bt)
  \right)
\end{equation}
which is retarded (advanced) for $x^+>0$ ($x^+<0$). Therefore, the
result~(\ref{h++sh}) confirms the delay-shift of sec.~\ref{s:sdts}, in a form
which is appropriate for the calculation of the real part of the action
$\Re \A$, or to the principal-value prescription of the $x^-,x^+$ propagator.

\begin{figure}[t]
  \centering
  \includegraphics[width=0.65\textwidth]{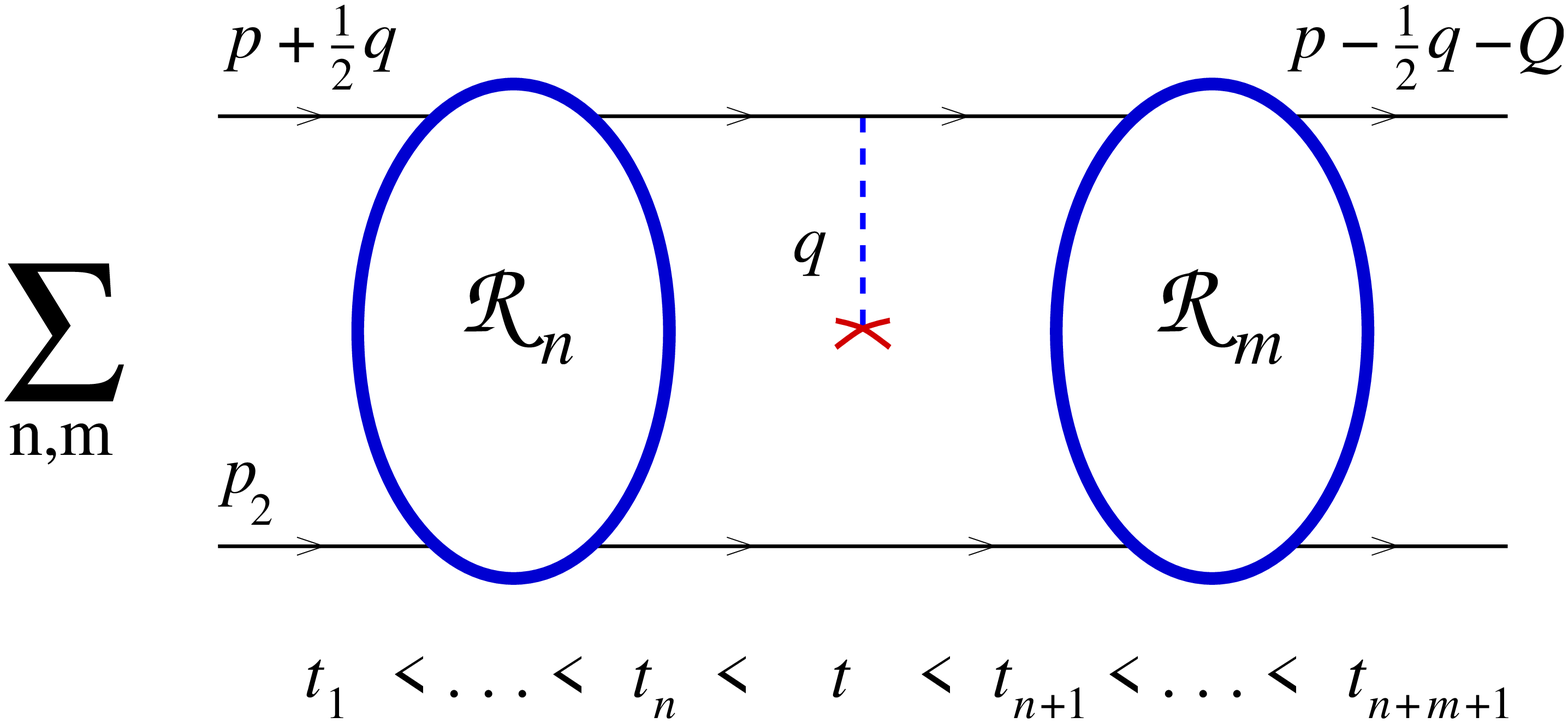}
  \Caption{Higher order eikonal diagrams with one insertion of the field
    $h_{--}$ (red cross). The blobs represent ladders of eikonal exchanges.}
  \label{f:2bolle}
\end{figure}

\begin{figure}[t]
  \centering
  \includegraphics[width=0.36\textwidth]{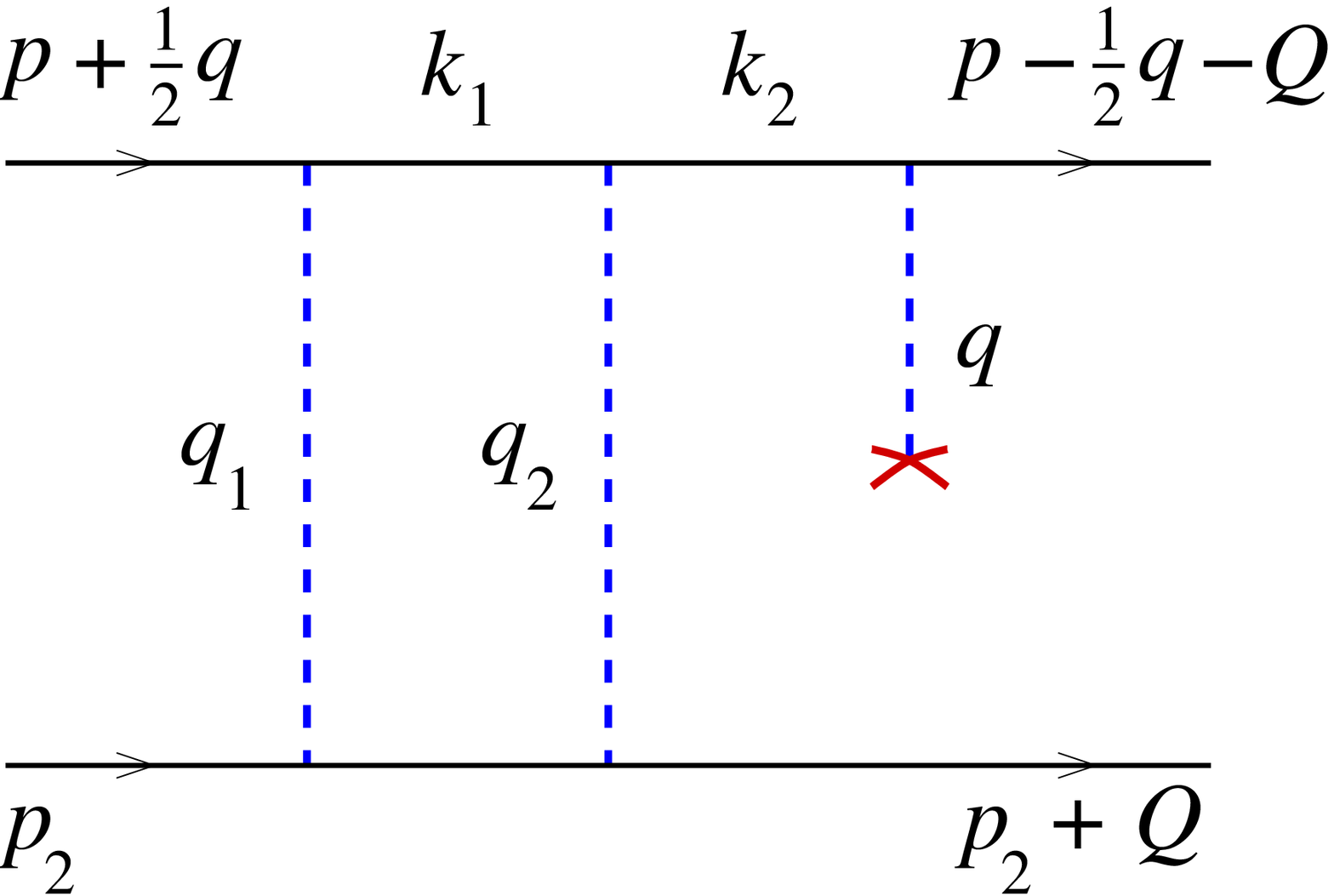}
  \Caption{Eikonal diagram with two graviton exchanges and one insertion of the
    field $h_{--}$ (red cross).}
  \label{f:2prop1x}
\end{figure}

The higher-order contributions to the shifted field are computed from their
definition in fig.~\ref{f:2bolle}. In order to do the explicit calculation, we
have to explain in more detail how the $x^+$-ordering just mentioned is generated
from the usual $t$-ordering in perturbation theory. We shall work off
energy-shell, but on mass-shell for the intermediate particles so that, in the
example of fig.~\ref{f:2prop1x}, $E_0 = |\vp| = |\vk_0|$, $E_1=|\vk_1|$, \dots,
$\bar{E}_0=|\vec{\bar{k}}_0|$. Therefore, to each eikonal exchange, we assign
the energy differences in the center of mass frame $\Delta_i=\Delta(\kt_i,\kt_{i-1})$, e.g.,
\begin{align}\label{Delta1}
  \Delta_1 &\equiv |\vk_1|+|\vk_1|-|\vk_0|-|\vk_0| = 2(|\vp-\vq_1|-|\vp|)
  \simeq 2 \frac{\vq_1^2-2\vp\cdot\vq_1}{2|\vp|} \\
 \label{Delta2}
  \Delta_2 &= 2(|\vk_1-\vq_2|-|\vk_1|)
  \simeq 2 \frac{\vq_2^2-2\vk_1\cdot\vq_2}{2|\vk_1|}
  = 2\frac{\vQ^2-2\vp\cdot\vQ}{2|\vp|}
\end{align}
and so on.%
\footnote{The approximate evaluation of the $\Delta_i$'s occurring in the last
  equation's line is sufficient to our purposes, because we shall need it around
  $\Delta_i=0$. The argument is therefore valid for the finite-angle kinematics
  too.
}
The ordered time integrations, up to time $t$ of the field, yield therefore the
factor
\begin{equation}\label{expiDelta}
  \frac{\exp[\ui(\Delta_1+\Delta_2)t]}{[\e+\ui(\Delta_1+\Delta_2)]\,
  [\e+\ui\Delta_1]} \to \frac{\exp[\ui(\Delta_1+\Delta_2)t]}{2
  (\e+\ui\Delta_2)(\e+\ui\Delta_1)} \;,
\end{equation}
where the last expression is obtained after $1\leftrightarrow 2$ symmetrization, because of
the integrations on $\Delta_1$, $\Delta_2$ induced by $\dif^3 q_1$,
$\dif^3 q_2$, with a factor which is symmetrical under $\vq_i$ permutations.

\begin{figure}[t]
  \centering
  \includegraphics[width=0.8\textwidth]{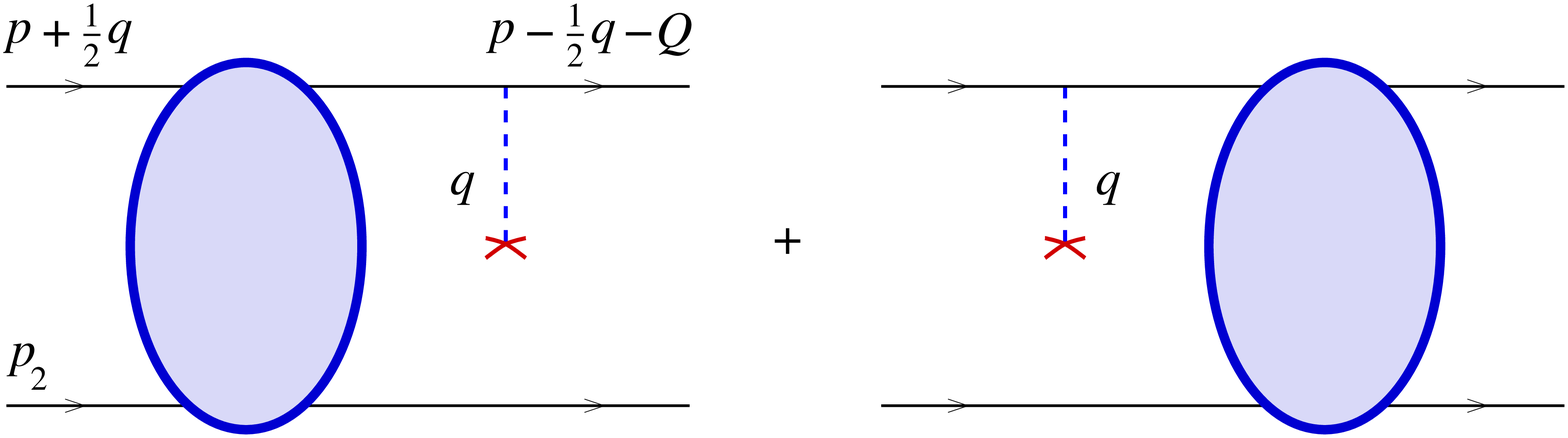}
  \Caption{Structure of the eikonal diagrams contributing to the field $h_{--}$.}
  \label{f:1bolla1x}
\end{figure}

Finally the factorized $\Delta_i$ integrations induced by the angular ones on the
$\vq_i$'s are done separately, by using the identity
\begin{equation}\label{DeltaIden}
  \frac1{2\pi}\int\dif\Delta \;\frac{\esp{\ui\Delta t}}{\e+\ui\Delta}
  = \Theta(t) + \ord{\esp{-\sqrt{s}|t|}}
\end{equation}
in which the integration contour is closed in the upper (lower) half-plane for
$t>0$ ($t<0$) by noting that corrections due to the large $\Delta$ contour are
exponentially small in $|\Delta_{\max}|=\ord{|\vq_i|}=\ord{\sqrt{s}}$ in our
finite angle kinematics. Similarly, the time and $\Delta$ integrations can be
done after the insertion, and yield
\begin{equation}\label{DeltaIden2}
  \frac1{2\pi}\int\dif\Delta \;\frac{\esp{\ui\Delta t}}{\e-\ui\Delta}
  \simeq \Theta(-t) \;.
\end{equation}
This means that eikonal echanges occurring both before and after the insertion
are suppressed,%
\footnote{Of course, a more refined calculation may introduce a spread
  $\Delta t \sim 1/\sqrt{s}$ in the $\Theta$-functions so that overlap of subsequent
  contributions is allowed for a limited time. The uncertainty so introduced in
  the time of shift is expected to be larger than $1/\sqrt{s}$ because of the
  number of coherent reinteractions, but in any case smaller than $R$, which
  would correspond to the maximal number $Gs/\hbar$ of leading eikonal scatterings.
}
because of $\Theta(t)\Theta(-t)=0$, so that the whole sum reduces to the
diagrams in fig.~\ref{f:1bolla1x}.

In order to compute them we first diagonalize the scattering amplitude in
impact parameter $\bt$ space and we then compute the inserted field by Fourier
transform in the $q^\mu$ variables, in order to yield its $x^-,x^+,\xt$
dependence. The non-trivial point about impact parameter transform is that it is
2-dimensional, while the integration variables $\vq_1$ and $\vQ-\vq_1$ (in, say
fig.~\ref{f:2prop1x}) are 3-dimensional in the finite angle
kinematics. Fortunately, the conditions $\Delta_1=0=\Delta_2$ fix the
3-components of both $\vq_1$ and $\vQ$, by determining the directions of the
respective Breit frames. We thus have the Bessel transform
\begin{equation}\label{besselTr}
  \frac12\int\frac{\dif^3 q_1}{2\pi^2 q_1^{\;2}}\; \delta\left(q_1\cos\alpha_1
  -\frac{q_1^2}{\sqrt{s}}\right) \; \frac{\dif^3 Q}{2\pi^2}\;
  \frac{\delta\left(Q\cos\beta_1-\frac{Q^2}{\sqrt{s}}\right) J_0(bQ)}{Q^2+q_1^2
  -2 Q q_1(\cos\alpha_1\cos\beta_1-\sin\alpha_1\sin\beta_1\cos\phi_1)}
\end{equation}
which, by the clever identity~(\ref{azimuthal}) yields the same result as the
2-dimensional azimuthal average
\begin{align}
  &\frac{\ui^2}{2}\int\frac{q_1\dif q_1\dif\phi_1}{2\pi^2 q_1^2}\;
  \frac{Q\, \dif Q\, \dif\phi_Q}{2\pi^2(q_1^2+Q^2-2Q q_1\cos\phi_1)}
  \esp{-\ui b Q \cos\phi_Q} \nonumber \\
  &= \frac{\ui^2}{2}\int\frac{\dif^2\qt_1\,\dif^2\Qt}{(2\pi^2)^2(\Qt-\qt_1)^2\qt_1^2}
  \esp{-\ui\bt\cdot(\Qt-\qt_1+\qt_1)} = \frac12\big(a_0(\bt)\big)^2
\end{align}
which is a 2-dimensional convolution diagonalized in $\bt$-space.

We thus obtain that the amplitude in front of the field insertion resums to the
exponentiated result
\begin{equation}\label{expRes}
  \sum_n \frac1{n!} \big(\ui 2\pi G s a_0(b)\big)^n = \esp{\ui 2\pi G s a_0(b)} = S(b,s)
\end{equation}
and cancels out with the factor $S^{-1}$ in eq.~(\ref{derFunz}).

We are left with the calculation of the $h_{--}$ field itself. Notice that in
fig.~\ref{f:1bolla1x}.b the insertion is on the incoming leg $\vp$, while in
fig.~\ref{f:1bolla1x}.a it occurs on the final leg $\vk=\vp-\vQ$. Since
$|\vp-\vQ|=|\vp|$ because of the mass-shell plus energy-conservation
constraints, the difference is just a rotation $\vk=\vp_R$ by the angle
$\sin\frac{\theta}{2} = Q/\sqrt{s}$. When inserted in the field propagator it
leads to a rotated variable%
\footnote{The $++$ projection should be rotated also with the $z$-axis in the
  direction of the deflected longitudinal direction.}
$\vx_R$ with a mass-shell condition%
\footnote{Strictly speaking, we should replace this mass-shell condition by the
  denominator $[(p-q)^2+\ui\e]^{-1}$. However, for $x^+ > 0$ the contour can be
  closed on the pole, with the same result, while for $x^+ < 0$ it
  vanishes. This remark confirms the ordering $\Theta(x^+)$  (see app.~\ref{a:hdi}).}
$(p-q)^2=-q^-(\sqrt{s}-q^+)-\qt^2=0$, and we get
\begin{equation}\label{da4a2}
  \int\frac{\dif^4 q}{(2\pi)^3}\,\frac{\esp{\ui\vq\cdot\vx_R}
  \esp{-\ui q_0 t} (2p^0-q^0)}{\vq^{\;2}-q_0^2}\delta\big((q-p)^2\big)
 = \int\frac{\dif^2\qt}{(2\pi)^2}\frac{\esp{\ui\qt\cdot\xt_R}
  \esp{-\ui\frac{q^+}{2}x^-_R}}{\qt^2} (p^+ - \half q^+)
  \frac{\dif q^+}{4\pi} \;,
\end{equation}
where we have neglected the exponent dependence on $q^- x^+$, when
$q^- = -\qt^2/(\sqrt{s}-q^+)$ is a very small variable if we make the natural
assumption $|\vq|\ll\sqrt{s}$, insuring that the field insertion does not modify
the scattering angle.

\begin{figure}[t]
  \centering
  \includegraphics[width=0.4\linewidth]{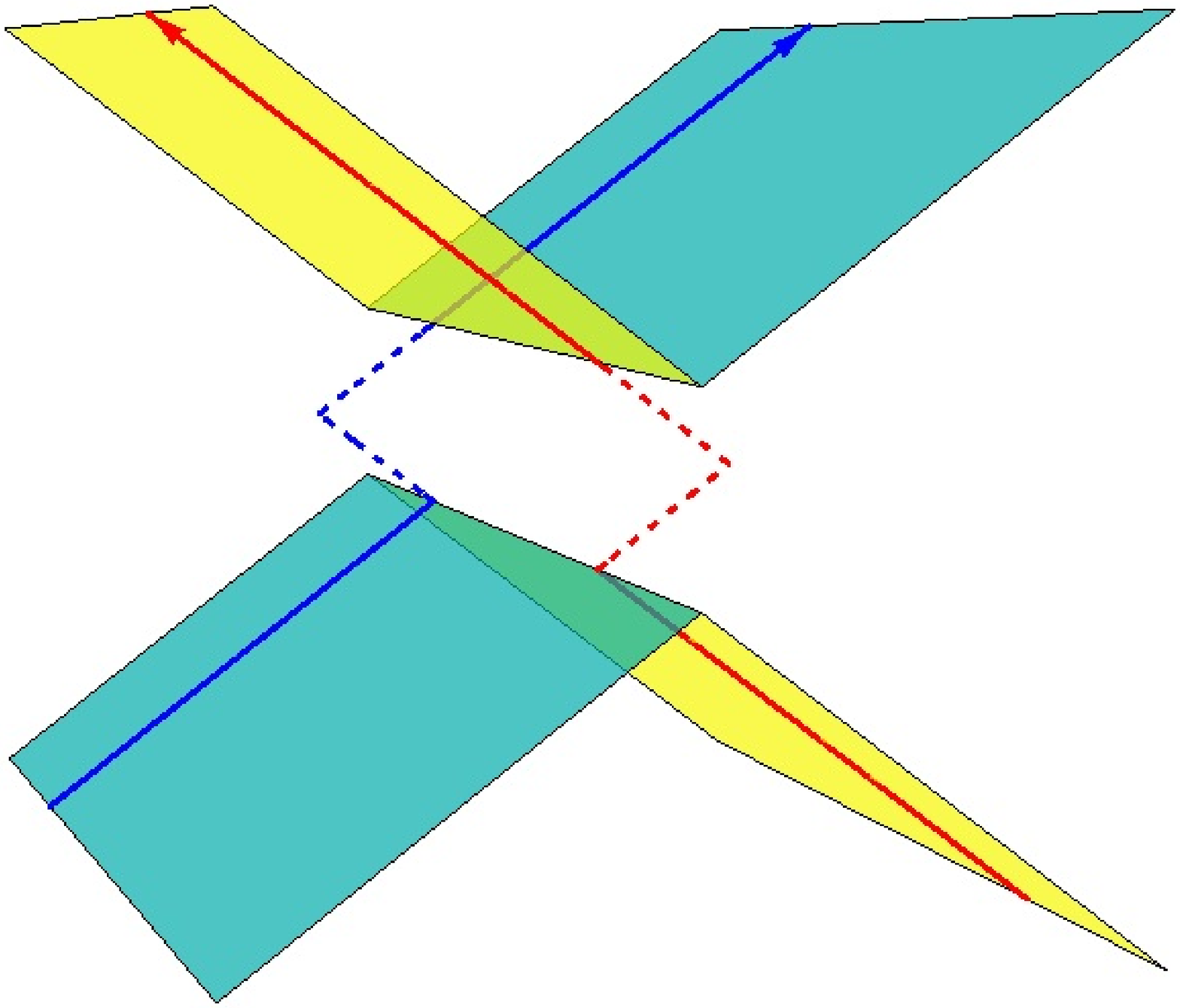}
  \Caption{Graphical representation of the shifted shock-waves in Minkowski
    space (only one transverse dimension is visible). The cyan (yellow)
    half-planes show the support of the field $h_{--}$ ($h_{++}$) generated by
    particle 1 (2), here represented by the blue (red) line.}
  \label{f:shiftedShocks}
\end{figure}

Finally, by taking the value $p^+=\sqrt{s}+\frac12 q^+$ of the initial energy as
we did in the beginning, we get for the field a factorized value proportional to
$\delta(x^-)$ which translates $\Theta(t)$ into $\Theta(x^+)$. The complete
resummation in fig.~\ref{f:2bolle} is thus summarized in the exponentiated form
\begin{align}
  h_{--}(x) &= S^{-1}\exp\left[2\pi G \sqrt{s} a_0(b)\left(\ui\sqrt{s}
 -\frac{\partial}{\partial x^-}\right)\Theta(x^+) + \left(\ui\sqrt{s}
  +\frac{\partial}{\partial x^-}\right)\Theta(-x^+)\right] \times \nonumber \\
 &\qquad\left\{2\pi R\left[\Theta(-x^+)\delta(x^-)a_0(\xt)+\Theta(x^+)\delta(x^-_R)a_0(\xt_R)
 \right]\right\} \nonumber \\[1mm]
 &= 2\pi R \left[ \delta\big(x^-_R-\pi R a_0(b)\big) a_0(\xt_R)\Theta(x^+)
  +\delta\big(x^- +\pi R a_0(b)\big) a_0(\xt)\Theta(-x^+)\right]\;. \label{hppR}
\end{align}

An identical calculation with the exchange $(+\leftrightarrow -)$ yields an
analogous result for the field $h_{++}(x)$. The overall situation is depicted in
fig.~\ref{f:shiftedShocks}, where we show the shifted (and slightly rotated)
shock waves, on which the source particles lie before and after the collision
(solid lines). The evolution of the particles during the collision actually
depends on the reference frame used to describe their motion. For instance, a
time-like observer that crosses first shock 2 (the one on the bottom-right)
describes the scattering in two steps: firstly the shock 2 drags shock 1 and
particle 1 by an amount $2\pi R \bar{a}(x)$ in the $x^-$ direction; then shock 1
drags shock 2 and particle 2 in the $x^+$ direction. In this frame particle 1
travels along the dashed blue line in fig.~\ref{f:shiftedShocks}.  A symmetric
description is given by an observer crossing first shock 1, which ``sees''
particle 2 moving along the dashed red line.

%%%%%%%%%%%%%%%%%%%%%%%%%%%%%%%%%%%%%%%%%%%%%%%%%%%%%%%%%%%%%%%%%%%%%%%%%%%%%%%%
\section{Rescattering corrections to the reduced-action\\ model\label{s:rcram}}
%%%%%%%%%%%%%%%%%%%%%%%%%%%%%%%%%%%%%%%%%%%%%%%%%%%%%%%%%%%%%%%%%%%%%%%%%%%%%%%%

%%%%%%%%%%%%%%%%%%%%%%%%%%%%%%%%%%%%%%%%%%%%%%%%%%%%%%%%%%%%%%%%%%%%%%%%%%%%%%%%
\subsection{Semiclassical field equations\label{s:sfe}}
%%%%%%%%%%%%%%%%%%%%%%%%%%%%%%%%%%%%%%%%%%%%%%%%%%%%%%%%%%%%%%%%%%%%%%%%%%%%%%%%

The transplanckian field equations in the ACV proposal~\cite{ACV90} were based
on two main groups of results. Firstly, the ACV investigation of string-gravity
showed that at transplanckian energies $Gs\gg 1$
($R\equiv 2G\sqrt{s} \gg 1/\sqrt{s}$) the eikonal representation --- with an
eikonal operator which is calculable in principle by expanding in $\lambda_s/b$
and $R/b$ --- yields a good representation of the scattering amplitude, and
incorporates both string- and strong-gravity effects.

Secondly, in the regime $b \gtrsim R \gg \lambda_s$ in which string effects are
supposed to be small, the irreducible eikonal diagrams are much in
correspondence with the effective action of Lipatov and
co-workers~\cite{Li91,KiSz95}, who calculated a Regge-graviton emission
vertex~\cite{Li82} which is the building block of the effective lagrangian used
in~\cite{ACV93}. Finally, the equations of motion of the latter --- by
neglecting rescattering terms --- were shown to yield a shock-wave solution for
the fields which is the basis for the reduced-action model investigated in
detail in later years~\cite{ACV07}.

Here, we have considered so far only the leading graviton-exchange kernel in the
eikonal, and we have improved the amplitude representation based on it, so as to
include a motion of the Breit-frame on which the exchange is defined. Next, we
want to reconsider the field equations in 4-dimensions --- including
rescattering --- and we shall provide a solution for the fields which fits very
well in our understanding of scattering developed so far, by adding corrections
which are of relative order $R^2/b^2$ and higher.

\begin{figure}[t]
  \centering
  \includegraphics[width=0.95\textwidth]{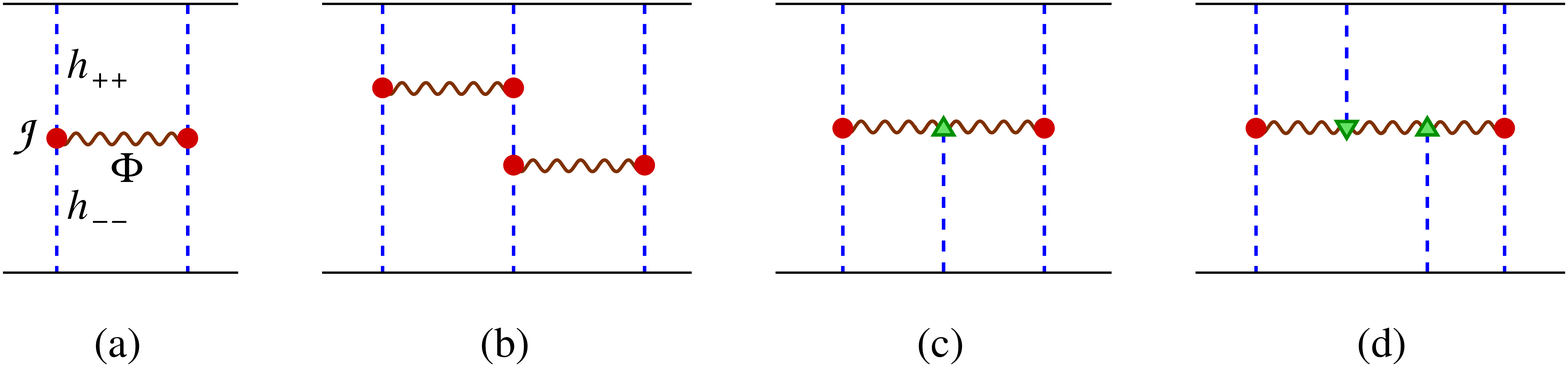}
  \Caption{Diagrams providing subleading contributions to the eikonal
    approximation: {\em(a)} the ``H-diagram'' representing the first correction of
    (relative) order $(R/b)^2$; {\em(b)} a ``multi-H'' diagram of order $(R/b)^4$
    involving only Lipatov's vertices (red disks); {\em(c)} the first diagram with a
    rescattering vertex (green triangle) is of order $(R/b)^3$, but actually
    vanishes on-shell; {\em(d)} the first nonvanishing rescattering diagram is of
    order $(R/b)^4$.}
  \label{f:sublDiagrams}
\end{figure}

In the effective action framework, the elastic $S$-matrix of the tree diagrams
in fig.~\ref{f:sublDiagrams} is given in terms of the classical solutions of the
lagrangian equation of motion as
\begin{align}\label{elasticS}
  S(b,s) &= \exp\left\{\frac{\ui}{\hbar} A(h_{\mathrm cl}^{\mu\nu})\right\} \\
 \label{action}
  A(\hti^{++},\hti^{--},\Phi) &= \int \dif^4 x\; (\L_0+\L_e+\L_r
  + T_{++} \hti^{++}+ T_{--} \hti^{--} ) \;,
\end{align}
where $\hti^{++}\equiv(2/\kappa)h_{--}$ and $\hti^{--}\equiv(2/\kappa)h_{++}$
are just rescaled versions of the longitudinal fields considered so far, $\Phi$
is related to the transverse field which is proportional to $h=\nabla^2 \Phi$,
and
\begin{equation}\label{enMomTens}
  T_{--} = \kappa E \delta(x^-)\delta^2(\xt) \;, \qquad
  T_{++} = \kappa E \delta(x^+)\delta^2(\xt-\bt)
\end{equation}
represents (up to an unconventional but convenient factor of $\kappa$) the
energy-momentum tensor of the colliding particles. The usual metric components
are given by~\cite{ACV07}
\begin{align}
  h_{\mu\nu}\dif x^\mu \dif x^\nu
  &\equiv \dif s^2 - \eta_{\mu\nu}\dif x^\mu \dif x^\nu \nonumber \\
  &= \frac{\kappa}{2}[\hti^{++}(\dif x^-)^2 + \hti^{--}(\dif x^+)^2]
  + \frac{\kappa}{4} \left[ \epsilon_{\mu\nu}^{TT} \gr^2\Re\Phi
  - \epsilon_{\mu\nu}^{LT} \gr^2\Im\Phi\right] \dif x^\mu \dif x^\nu \;,
 \label{hmunudxdx}
\end{align}
where $\gr^2$ denotes the transverse laplacian and the $\epsilon$'s denote
graviton polarizations to be specified below.

The lagrangian is conveniently written in terms of real light-cone variables
$x^{\pm}$, $\partial_{\pm}\equiv\partial/\partial x^\pm$ and a complex transverse variable
$z\equiv x^1+\ui x^2$, $\partial\equiv\partial/\partial z =
\half(\partial_1-\ui\partial_2)$. It consists of a kinetic term
\begin{equation}\label{kinTerm}
  \L_0 = -\partial^*\hti^{++}\partial \hti^{--}
  +4\partial_+\dsq\Phi \partial_-\dq\Phi^* \;,
\end{equation}
where the longitudinal fields have a mostly transverse propagator and the
(complex) $\Phi$ field a mostly longitudinal one, of a graviton emission term
\begin{equation}\label{emissionTerm}
  \L_e = \kappa(\J\dmq\Phi^*+\J^*\dmq\Phi) \;, \qquad
  \dmq \J = \dsq\hti^{++}\dq\hti^{--}
  - \dmq \hti^{++} \dmq \hti^{--}
\end{equation}
incorporating Lipatov's vertex~\cite{Li82}, and finally, of a rescattering term
\begin{equation}\label{rescattering}
  \L_r = \kappa (\hti^{++}\dsq\Phi^*\partial_+^2\dq\Phi
  + \hti^{--}\dq\Phi^*\partial_-^2\dsq\Phi) \quad+\quad
  (\partial_+ \leftrightarrow \partial_-)\;,
\end{equation}
which is supposed to take into account the rescattering diagrams of
fig.~\ref{f:sublDiagrams}.c,d. This term is quadratic in $\Phi$, and is likely to
play a role when the latter is large.

Notice in particular in eq.~(\ref{emissionTerm}) the current $\J$, which
describes transverse graviton emission~\cite{Li91} in the so-called
H-diagram~\cite{ACV90} and is non-local because of the inverse laplacian needed to
find it. For this reason it has been convenient to introduce the complex field
$\Phi$, which corresponds to two graviton polarizations. The detailed
computation shows that
\begin{equation}\label{hzz}
  h_{z\bar{z}} = \kappa \dmq\Phi\;, \qquad
  h_{zz} = \kappa \dq\Phi \;, \qquad h_{\bar{z}\bar{z}} = \kappa \dsq\Phi
\end{equation}
and, more generally, using a notation with real-valued indices, that
\begin{equation} \label{hmunu}
  h_{\mu\nu} = \kappa \Re\left(-\eta^*_\mu\eta^*_\nu \dmq\Phi \right) \;,
\end{equation}
where
\begin{equation}\label{epsLT}
  \eta^\mu(k) \equiv \e_L^\mu+\ui\e_T^\mu = \Big(\frac{k^3}{|\kt|},\ui\bs{\e},
  \frac{k^0}{|\kt|} \Big)
\end{equation}
so that we have the polarizations
\begin{equation}\label{polarizations}
  \e_{TT}^{\mu\nu} = (\e_T^\mu \e_T^\nu - \e_L^\mu \e_L^\nu)\;, \quad
  \e_{LT}^{\mu\nu} = (\e_L^\mu \e_T^\nu + \e_T^\mu \e_L^\nu)\;, \quad
  \e_{\mu\nu}^i \e_j^{\mu\nu} = 2\delta^i_j \qquad
  (i,j = TT,LT) \;.
\end{equation}

The resulting equations of motion are given by
\begin{subequations}\label{eqs}
  \begin{align}
    &\dmq\hti^{++} + \kappa \left[\dq(\Phi^*\dsq\hti^{++}
      +\dsq(\Phi\dq\hti^{++})-\dmq((\Phi+\Phi^*)\dmq\hti^{++}) \right] +
    \frac12\kappa\sqrt{s}\delta^2(\xt)\delta(x^-) \nonumber \\
    &\qquad = -\kappa\left(\dq\Phi^*\dq_-\dsq\Phi
      \;+\;(\partial\leftrightarrow\partial^*) \right)
    \label{eqh} \\
    &4\partial_+\partial_- |\partial|^4\Phi -\kappa \left[
      \dsq\hti^{++}\dq\hti^{--} - \dmq\hti^{++}\dmq\hti^{--}\right] \nonumber \\
    &\qquad= \kappa  \left[ \dsq\left(\hti^{++}\dq_+\dq\Phi\right) +
      \dq\left(\hti^{--}\dq_-\dsq\Phi\right)
      \;+\;(\partial\leftrightarrow\partial^*)\right] \;,
    \label{eqphi}
  \end{align}
\end{subequations}
where the rescattering terms are set on the r.h.s., with similar equations for
$\hti^{--}$ and $\Phi^*$.

Notice that in eqs.~(\ref{eqs}) the impinging particles are represented by the
frozen external sources $T^{++}$ and $T^{--}$ written in eq.~(\ref{enMomTens}),
which pertain to the collinear kinematics used before. According to this
small-angle approach, scattering comes out because of a non-trivial action, but
is not taken into account in the energy-momentum of the particles. Shortly, we
shall take a somewhat different attitude.

%%%%%%%%%%%%%%%%%%%%%%%%%%%%%%%%%%%%%%%%%%%%%%%%%%%%%%%%%%%%%%%%%%%%%%%%%%%%%%%%
\subsection{The reduced-action without rescattering\label{s:rawr}}
%%%%%%%%%%%%%%%%%%%%%%%%%%%%%%%%%%%%%%%%%%%%%%%%%%%%%%%%%%%%%%%%%%%%%%%%%%%%%%%%

We start recalling~\cite{ACV93} that, if the rescattering terms are dropped,
eqs.~(\ref{eqs}) admit a shock-wave solution of the form
\begin{subequations}\label{separation}
  \begin{align}
    4 \hti_{--} = \hti^{++} &= \kappa\sqrt{s}\delta(x^-) a(\xt) \label{sep_h++} \\
    4 \hti_{++} = \hti^{--} &= \kappa\sqrt{s}\delta(x^+) \bar{a}(\xt) \label{sep_h--} \\
    \Phi &= \frac{\kappa^3 s}{2}\,\frac12\Theta(x^+ x^-) \phi(\xt) \;,
    \label{sep_phi}
  \end{align}
\end{subequations}
where the metric component
$h_{--} = (\kappa/2) \hti^{++} = 2\pi R a(\xt)\delta(x^-)$ are of AS
type~\cite{AiSe} while the transverse field $\Phi$ has support inside the whole
{\em light-wedges} $x^+ x^- \geq 0$. The latter propagation, of retarded plus
advanced type, corresponds to the principal value part of the Feynman
propagators, as is appropriate for the real part of the amplitude.

The field equations (\ref{eqs}) induce a set of differential equations on $a$,
$\bar{a}$ and $\phi$, which can be solved on the basis of proper boundary
conditions which make them regular, i.e., ultraviolet safe. Furthermore, by
replacing the expressions~(\ref{separation}) into~(\ref{eqs}) and omitting the
rescattering term, we obtain the ACV equations of motion and the corresponding
reduced action
\begin{align}
  \A_R = 2\pi Gs\Big\{a(\bt) + \bar{a}(0) + \int \dif^2\xt\;&\Big[
  -\frac12\gr a \cdot \gr\bar{a} - \frac12(\pi R)^2 |\gr^2\phi|^2 \nonumber\\
  &+(2\pi R)^2 \big(\phi^* (\dsq a\dq\bar{a}-\dmq a \dmq\bar{a}) + \text{h.c.}
  \big) \Big] \Big\} \;, \label{redA}
\end{align}
where we have freely performed some integration by parts, assuming a smooth
enough ultraviolet behaviour of the solutions, and we have replaced the current
$\J$ by its reduced counterpart $\Hc$:
\begin{equation}\label{defHc}
  \J \equiv \frac{\kappa^2 s}{2}\delta(x^-)\delta(x^+)\Hc \qquad\imp\qquad
  \Hc = \frac{2}{\dmq}(\dsq a\dq\bar{a}-\dmq a\dmq \bar{a}) \;.
\end{equation}
We state below some of the properties of the RAM solutions~\cite{ACV07} in the
particularly simple case of axisymmetric sources like particle-ring scattering
--- which is relevant for the azimuthal averaged particle-particle case.

By assuming the real fields $a(r^2)$, $\bar{a}(r^2)$ and the complex field
$\phi(r^2)$ to be functions of $r^2\equiv|\xt|^2$ only, and by calling $s(r^2)$,
$\bar{s}(r^2)$ the axisymmetric sources, we soon realize that the action $\A_R$
can be recast in one-dimensional form. This is because the current $\Hc$ defined
by eq.~(\ref{defHc}) is simply obtained in this case by
setting $\dot\Hc = -2\dot{a}\dot{\bar{a}}$ --- where the dot denotes the
$r^2$-derivative --- and is therefore real and axisymmetric also. As a
consequence, the interaction term involves $\Re\phi$ only and is proportional to
the $a\bar{a}$ kinetic term as follows
\begin{equation}\label{1dimAction}
  \A_R = 2\pi^2 G s \int\dif r^2\left[\bar{s}a+s\bar{a} - 2\rho \dot{a}
  \dot{\bar{a}} - \frac{2}{(2\pi R)^2}(1-\dot{\rho})^2\right]
  \qquad \left(\cdot \equiv \frac{\dif}{\dif r^2} \right) \;.
\end{equation}
Here we have replaced $\phi$ by the auxiliary field $\rho(r^2)$ --- a sort of
renormalized squared distance --- defined by
\begin{equation}\label{rho}
  \rho \equiv r^2 [1-(2\pi R)^2\dot\phi] \;, \qquad
  h \equiv \gr^2\phi = 4\frac{\dif}{\dif r^2}(r^2\dot\phi)
  = \frac{1}{(\pi R)^2}(1-\dot\rho) \;,
\end{equation}
which incorporates the $\phi$-$a$-$\bar{a}$ interaction. Furthermore, the field
$\phi$ is now taken to be real-valued, describing the $TT$ polarization
only. The external axisymmetric sources $s$ and $\bar s$ are able to
approximately describe the particle-particle case by setting
$s(r^2)=\delta(r^2)/\pi$, $\bar{s}(r^2)=\delta(r^2-b^2)/\pi$, where the
azimuthal ACV averaging procedure is assumed.

The equations of motion, derived from the reduced-action~(\ref{1dimAction}) and
specialized to the case of particles at impact parameter $b$ (with the
axisymmetric sources just quoted), provide the profile functions
\begin{equation}\label{eoma}
  \dot{a} = -\frac1{2\pi\rho} \;, \qquad
  \dot{\bar{a}} = -\frac1{2\pi\rho}\Theta(r^2-b^2) \;,
\end{equation}
from the analogue of eq.~(\ref{eqh}), and the $\rho$-field (or $\phi$-field)
\begin{equation}\label{eomrho}
  \ddot{\rho} = \frac1{2\rho^2}\Theta(r^2-b^2) \;, \qquad
  \dot{\rho}^2+\frac1{\rho} = 1 \qquad (r > b)
\end{equation}
from the analogue of eq.~(\ref{eqphi}). Eq.~(\ref{eomrho}) shows a ``Coulomb''
potential in $\rho$-space, which is repulsive for $\rho>0$, acts for $r>b$ and
plays an important role in the tunneling phenomenon~\cite{CiCoFa}. By replacing
the equation of motion~(\ref{eoma}) into eq.~(\ref{1dimAction}), the reduced
action can be expressed in terms of the $\rho$ field only, and takes the simple
form
\begin{equation}\label{rhoAction}
  \A_R = -Gs\int\dif r^2
  \left[\frac1{R^2}(1-\dot\rho)^2-\frac1{\rho}\Theta(r^2-b^2)\right]
  \equiv -\int_0^\infty \dif r^2 \; \L(\rho,\dot\rho,r^2) \;,
\end{equation}
which is the one we shall consider at quantum level in the following.

The effective metric generated by the axisymmetric fields $\rho$, $a$ and
$\bar{a}$ was calculated~\cite{ACV07} on the basis of the complete form of the
shock-wave~(\ref{separation}) and is given by
\begin{align}
 \dif s^2 &= -\dif x^+\dif x^- 
 \left[1-\half \Theta(x^{+}x^{-})(1-\dot\rho)\right]
 \nonumber \\
 &+(\dif x^+)^2 \delta(x^{+}) \left[
   2\pi R \bar{a}(r^2) -\fourth(1-\dot\rho) |x^{-}|
 \right]
 \nonumber \\
 &+(\dif x^-)^2 \delta(x^{-}) \left[
   2\pi R a(r^2) -\fourth(1-\dot\rho) |x^{+}|
 \right]
 \nonumber \\
 &+ \dif r^2\left[1+2(\pi R)^2\Theta(x^{+}x^{-})\dot{\phi} \right]
 +\dif\theta^2\,r^2 \left[1+2(\pi R)^2\Theta(x^{+}x^{-})(\dot{\phi}
  +2r^2\ddot{\phi})\right] \;.
 \label{metr}
\end{align}
This metric is dynamically generated and may be regular or singular at short
distances, depending on the behaviour of the field solutions themselves. It is
not fully consistent, however, since it does not take into account the
longitudinal shifts that the fields cause to each other. In the following
sections we shall compute such shifts at subleading level and we shall present
the improved expression of the self-consistent metric.

%%%%%%%%%%%%%%%%%%%%%%%%%%%%%%%%%%%%%%%%%%%%%%%%%%%%%%%%%%%%%%%%%%%%%%%%%%%%%%%%%
\subsection{The H-diagram: scattering angle and shifts\label{s:Hd}}
%%%%%%%%%%%%%%%%%%%%%%%%%%%%%%%%%%%%%%%%%%%%%%%%%%%%%%%%%%%%%%%%%%%%%%%%%%%%%%%%%

The first nontrivial use of the reduced action~(\ref{redA}) is the calculation
of the first order correction to the Einstein deflection~\cite{ACV90} which is
due to the H-diagram contribution to the action (fig.~\ref{f:sublDiagrams}.a). The latter is
obtained by expanding the action~(\ref{redA}) and the corresponding equations of
motion (quoted in eqs.~(\ref{eoma},\ref{eomrho}) for the axisymmetric case) in the
parameter $R^2/b^2$ where $R\equiv 2G\sqrt{s}$ is the gravitational radius and
$b$ is the impact parameter conjugated to the transverse momentum $Q$ (and
related to the true impact parameter $b_0$ by eq.~(\ref{b0})).

By expanding the equations of motion we obtain first order corrections to the
profile function
\begin{equation}\label{aExp}
  a(\xt) = a_0(\xt) + a_1(\xt) + \cdots
\end{equation}
and to the transverse field
\begin{equation}\label{hExp}
  h(\xt) = \gr^2\phi(\xt) = h_0(\xt) + \cdots = \frac{2}{\dmq}
  \left(\dsq a_0\dq\bar{a}_0-\dmq a_0 \dmq\bar{a}_0\right) + \cdots
\end{equation}
as consequence of the Lipatov's vertex $\J$ in eq.~(\ref{emissionTerm}). It is
then straighforward to obtain for the action~\cite{ACV07}
\begin{align}
  a(b) &= a_0(b) + 2 a_H(b) + \cdots \\
  \A_R &= 2 \pi G s \left[ a_0(b) + a_H(b) + \cdots \right] \;,
\end{align}
where
\begin{equation}\label{a1}
  a_0(\xt) = \frac{1}{2\pi}\log\frac{L^2}{|\xt|^2} \;, \qquad
  a_1(b) = 2 a_H(b)
\end{equation}
and the H-diagram contribution is
\begin{equation}\label{Hdiagram}
  a_H(b) = \frac12 (\pi R)^2 \int\dif^2\xt\; |h_0(\xt)|^2 \;.
\end{equation}

In the following, we specialize the expression~(\ref{Hdiagram}) to the
axisymmetric case%
\footnote{This allows us to avoid an infrared divergence (present in
  eq.~(\ref{Hdiagram})) which is due to the $\e_{LT}$ polarization --- and is to
be subtracted out by the exponentiation procedure of ref~\cite{ACV90}.}
in which, according to eq.~(\ref{rho}),
\begin{equation}\label{hAxis}
  h(r^2) = \frac1{(\pi R)^2} \big(1-\dot\rho(r^2)\big)
\end{equation}
and --- by the equations of motion~(\ref{eomrho}) --- we obtain the result
\begin{equation}\label{aH}
  a_H = \frac1{2\pi R^2} \int_0^\infty \dif r^2 \; (1-\dot\rho)^2
  = \frac{R^2}{4\pi b^2}
\end{equation}
as contribution of the $TT$ polarization only. Here we have used the regular
solution of eq.~(\ref{eomrho})
\begin{equation}\label{regsol}
  \rho(\tau) = t_b \tau \Theta(b^2-\tau)+R^2\cosh^2\chi(\tau)\Theta(\tau-b^2)
\end{equation}
where $\tau-b^2 = R^2(\chi+\sinh\chi\cosh\chi-\chi_b -\sinh\chi_b\cosh\chi_b)$
and $t_b\equiv \tanh\chi_b$ is the largest real solution of eq.~(\ref{tbcond}).

The complete result, including the $LT$ polarization would be~\cite{ACV90}
\begin{equation}\label{ARH}
  \A_R = 2\pi Gs(a_0 + a_H + \cdots) = Gs \left(\log\frac{L^2}{b^2}
  + \frac{R^2}{b^2} + \cdots \right)
\end{equation}
and provides --- by stationarity of the eikonal phase --- the scattering angle
\begin{equation}\label{scatang}
  \sin\frac{\theta_{\mathrm cl}}{2}
  = -\frac1{\sqrt{s}} \frac{\partial}{\partial b} \A_R
  = \frac{R}{b}\left[1+\left(\frac{R}{b}\right)^2 + \cdots\right] \;.
\end{equation}

\begin{figure}[t]
  \centering
  \includegraphics[width=0.9\textwidth]{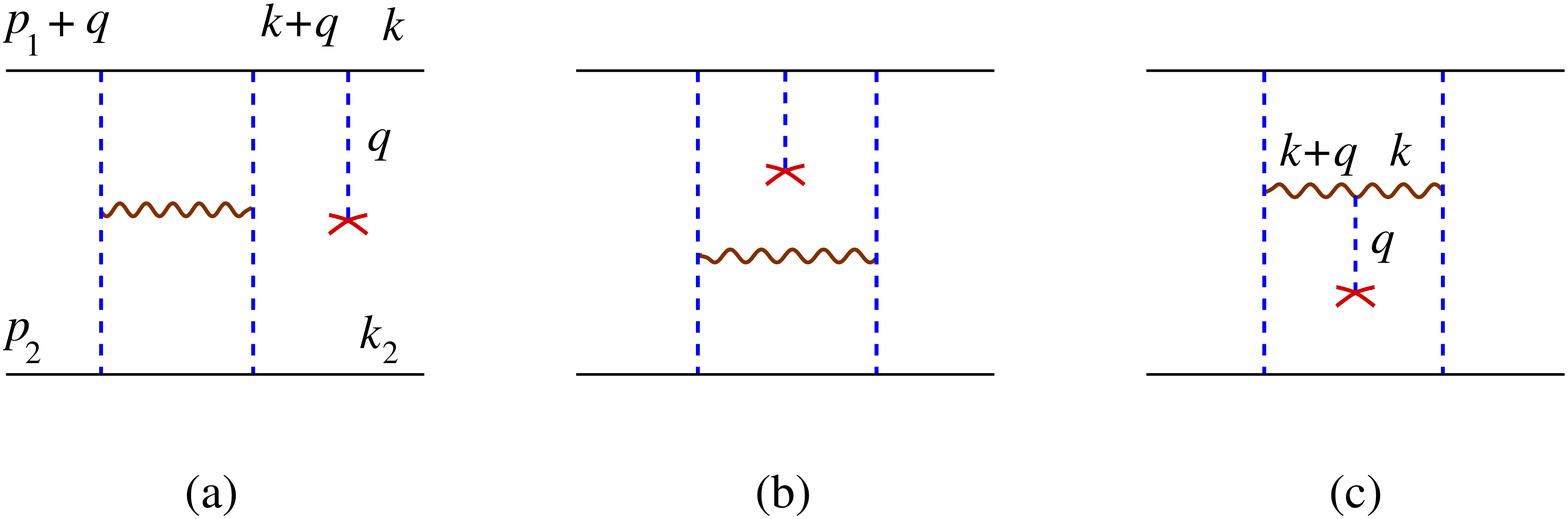}
  \Caption{Insertions of the field $h_{--}$ (red cross) on the H-diagram.}
  \label{f:HdiagramInsertions}
\end{figure}

The recollection above suggests to generalize the calculation of the shifted
field $h_{--}$ in sec.~\ref{s:ssw} to the next order in the $R/b$ expansion, by
including the H-diagram in the eikonal. In so doing we find 3 kinds of
insertions of the $h_{--}$ source, illustrated in
fig.~\ref{f:HdiagramInsertions}: {\it(a)} insertions on the on-shell
propagators, which can be done in a similar way as we did the leading one;
{\it(b)} insertions on the particle propagators of the H-dagram itself, and
{\it(c)} insertions on the emitted transverse field $h=\gr^2\phi$. The latter
are new and involve the rescattering vertex, which is omitted in the RAM
calculation and will be included in the next subsection.

Insertions of type {\it(a)} above add up to the leading shift $a_0(b)$ and are
done by replacing $p^+$ with $p^+ + q^+$ in the corresponding energy
charge.%
\footnote{We refer here to the overall shift $2\Delta(\xt,b)$ between past and
  future. Note that in the ACV equations framework the profile function $a(\xt)$
  and the shift $\Delta(\xt,b)$ occur together in the product $D(\xt,b)$
  (eq.~(\ref{eqD})), while in the insertion formalism
  $\Delta(\xt,b)\simeq\Delta(b)$ is probed by the leading profile $a_0(\xt)$
  only, which dominates for $|\xt|\gg b$.}
However, $Gs\, a_H \sim G^3 s^2$ contains two $p^+$ factors and thus
acquires a factor of 2, as follows
\begin{align}
  \A_R &\to 2\pi Gs\left[a_0\left(1+\frac{q^+}{\sqrt{s}}\right)
  + a_H \left(1+\frac{q^+}{\sqrt{s}}\right)^2 \right] \nonumber \\
 &= 2\pi Gs (a_0 + a_H) + 2\pi R(a_0 + 2 a_H)\ui\frac{\partial}{\partial x^-}
 + \cdots \;.
\end{align}
The outcome is that the overall shift $2\Delta$ is proportional to
$\bar{a}(0) = a(b)$, as given by the perturbative formula~(\ref{h++sh}):
\begin{equation}\label{DeltaH}
  2\Delta = 2\pi R (a_0+2a_H+\cdots) = 2\pi R\, a(b)
  = \frac{\partial}{\partial\sqrt{s}}\A_R(s,b)
\end{equation}
and thus it extends to the H-diagram the relationship between shift and
energy-derivative of the action suggested in sec.~\ref{s:ilepp}.

%%%%%%%%%%%%%%%%%%%%%%%%%%%%%%%%%%%%%%%%%%%%%%%%%%%%%%%%%%%%%%%%%%%%%%%%%%%%%%%%%
\subsection{Rescattering solutions with shifted fields\label{s:rssf}}
%%%%%%%%%%%%%%%%%%%%%%%%%%%%%%%%%%%%%%%%%%%%%%%%%%%%%%%%%%%%%%%%%%%%%%%%%%%%%%%%%

Consider now the full field equations, starting from~(\ref{eqh}). While, by
replacing in the l.h.s.\ the shock-wave~(\ref{separation}), we generate terms
proportional to $\kappa\sqrt{s}\delta(x^-)$, the rescattering term in the
r.h.s.\ yields instead the structure
\begin{equation}\label{ddeltaStruc}
  -\kappa\sqrt{s} (2\pi R)^3 \delta'(x^-) \half \sgn(x^+) |\dsq\phi|^2 \;,
\end{equation}
which suggests an $x^-$ translation, i.e., a shift similar to those just found
at leading and H-diagram level with the eikonal insertions. Therefore, we make
the Ansatz
\begin{subequations}\label{shifts}
  \begin{align}
    \hti^{++} &= \kappa\sqrt{s}\delta\big(x^- -\pi R\Delta(\xt) \sgn(x^+)\big)
    a(\xt) \label{hppShift} \\
    \hti^{--} &= \kappa\sqrt{s}\delta\big(x^+ -\pi R\bar\Delta(\xt)
    \sgn(x^-)\big)
    \bar{a}(\xt) \label{hmmShift} \\
    % \Phi &= \frac{\kappa^3 s}{2}\phi(\xt) \left[ \Theta(x^- -\pi
    %   R\Delta_\phi)\Theta(x^+ -\pi R\bar\Delta_\phi) +\Theta(-x^- -\pi
    %   R\Delta_\phi)\Theta(-x^+ -\pi R\bar\Delta_\phi)\right] \;,
    % \\
    \Phi &= \frac{\kappa^3 s}{2}\phi(\xt) \left[ \Theta(x^- -\pi
      R\Delta_\phi)\Theta(x^+ -\pi R\bar\Delta_\phi) +(x^\pm \to -x^\pm)\right]
    \;,
    \label{phiShift}
  \end{align}
\end{subequations}
where $\Delta,\bar\Delta,\Delta_\phi,\bar\Delta_\phi(\xt,b)$ are shift variables to be determined. We take a similar
Ansatz for energy-momentum too, because we know from last section the latter is
really shifted at leading level. By then formally expanding the
$\delta$-function~(\ref{hppShift}) in $R/x^-$, we get two equations from the zeros
of the $\delta$ and $\delta'$ contributions.

The first equation is simply the reduced-action model (RAM) equation for $a$,
which reads
\begin{equation}\label{eqa}
  \dmq a + \frac{2\pi R}{2} \left[ \dq(\phi^*\dsq a)-\dmq(\phi^*\dmq a) +
    \text{c.c.} \right] +\half\delta^2(\xt) = 0
\end{equation}
and reduces, in the axisymmetric case and for real-valued $\phi$, to
\begin{equation}\label{eqarho}
  \frac{\dif}{\dif r^2}\big[\dot a\, r^2\big(1-(2\pi R)^2 \dot\phi \big) \big]
  = \frac{\dif}{\dif r^2}\big[\dot a\rho] = -\frac1{2\pi}\delta(r^2) \;.
\end{equation}
The latter is equivalent to eq.~(\ref{eomrho}) and yields the solution
\begin{equation}\label{asol}
  a(r^2) = \frac1{2\pi}\int_{r^2}^{L^2} \frac{\dif r'{}^2}{\rho(r'{}^2)} \;,
\end{equation}
where we have introduced the infrared cutoff parameter $L$ by setting
$a(L^2)=0$. Since by eqs.~(\ref{regsol}) above~\cite{ACV07}
\begin{equation}\label{rhoasy}
  \rho(r^2) \simeq
  \begin{cases}
    r^2-R^2\log\frac{r^2}{4b^2} &(r^2 \gg b^2) \\
    t_b \, r^2 &(r^2 \leq b^2)
  \end{cases}
\end{equation}
$a(r^2)$ approaches the leading value for $b\gg R$
\begin{equation}\label{aasy}
  a(r^2) \simeq a_0(r^2) = \frac1{2\pi}\log\frac{L^2}{r^2} + \cdots
\end{equation}
and diverges logarithmically for $r^2\to0$.

The second equation (from the vanishing of the $\delta'$-coefficient) involves
the rescattering terms and is supposed to determine the shift parameter
$\Delta$, which occurs in the combination $D(\xt)\equiv\Delta(\xt)a(\xt)$, as
follows
\begin{equation}\label{eqD}
  \dmq D + \frac{(2\pi R)^2}{2}\left[\dq(\phi^*\dsq D)-\dmq(\phi^*\dmq D)
  +\text{c.c.} \right]
 = -\frac{\Delta_0}{2}\delta^2(\xt)+(2\pi R)^2 |\dsq\phi|^2 \;,
\end{equation}
where we have set the rescattering terms on the r.h.s.\ and we notice in
particular the energy-momentum shift which, according to sec.~\ref{s:ssw}, has
the contribution
\begin{equation}\label{Delta0}
  \Delta_0 = \bar{a}(0) = a(b) \;.
\end{equation}
emerging from the eikonal insertions up to $R^2/b^2$ accuracy.
In the axisymmetric limit, eq.~(\ref{eqD}) takes the simplified form
\begin{align}\label{eqrhoD}
  \frac{\dif}{\dif r^2}(\rho\dot D) &= -\frac{\Delta_0}{2\pi}\delta(r^2) +
  \Jr(r^2) \\
  \Jr(r^2) &\equiv \frac{\Theta(r^2-b^2)}{(2\pi R)^2} \left(\frac{\rho}{r^2}
 -\dot\rho\right)^2 = (2\pi R)^2(r^2\ddot\phi)^2 \;,
\end{align}
where we note that the rescattering source $\Jr(r^2)$ vanishes for
$r^2\leq b^2$, as a consequence of the equations of motion~(\ref{rhoasy}).

It is then straightforward to find a solution for $\Delta(r^2)$ which is regular
everywhere, in the form
\begin{equation}\label{solDelta}
  \Delta(r^2) =
  \begin{cases}
    \displaystyle
    \Delta(\infty) + \frac1{a(r^2)}\int_{r^2}^{L^2}\frac{\dif r'{}^2}{\rho(r'{}^2)}
    \int_{r'{}^2}^\infty\dif r''{}^2\,\Jr(r''{}^2) &(r^2\geq b^2) \\[3mm]
    \displaystyle
    \Delta_0 - \frac1{a(r^2)}\left[2\pi a(b) \int_{b^2}^\infty
      \dif r'{}^2\,\Jr(r'{}^2) - \int_{b^2}^{L^2}\frac{\dif r'{}^2}{\rho(r'{}^2)}
    \int_{r'{}^2}^\infty\dif r''{}^2\Jr(r''{}^2) \right] &(r^2\leq b^2)
  \end{cases}
\end{equation}
where
\begin{equation}\label{diffDelta}
  \Delta(\infty)=\Delta_0-2\pi\int_{b^2}^\infty \dif r^2 \; \Jr(r^2)
  = a(b)-\int_0^\infty\frac{\dif r^2}{2\pi R^2}\;(1-\dot\rho)^2 \;.
\end{equation}
In other words, the shift parameter of the longitudinal field takes over the
(constant) leading value $a(b)$ from the energy-momentum tensor and adds an
$r^2$-dependent term of relative order $R^2/b^2$ which, for $r^2\gg b^2$, is
related to the kinetic term of the RAM action
\begin{equation}\label{Akin}
  \A_R
  = 2\pi Gs\left[a(b)-\int_0^\infty
  \frac{\dif r^2}{2\pi R^2} (1-\dot\rho)^2 \right]
  \equiv 2\pi Gs \, \As(b) \;.
\end{equation}
This means that, while the RAM action is sufficient in order to describe the
scattering parameters at relative order $R^2/b^2$ (H diagram), the rescattering
terms are needed in order to describe the form of the fields, like shock-wave
shifts, and thus the metric properties at a comparable level of accuracy.

It is amusing to check the result (\ref{solDelta}) --- which is based on the
rescattering equations and on the energy momentum shift in eq.~(\ref{eqD}) ---
by using the direct insertions {\it(b)} and {\it(c)} on the H-diagram mentioned
before, and depicted in fig.~\ref{f:HdiagramInsertions}. By keeping track of the
flow of energy charges we calculate the above insertions on the imaginary parts
(app.~\ref{a:hdi}) and we find that diagram {\it(b)} is already counted by
Lipatov's vertices, while diagram {\it(c)} contributes the absorptive part
\begin{equation}\label{Hc}
  \kappa \sqrt{s} \, Gs \frac{R^2}{2} \int\frac{\dif^2 \qt}{2\pi^2 \qt^2}\;
  \esp{\ui\qt\cdot\xt} \int\dif^2 \kt\;\hti(\kt)\hti^*(\kt+\qt)
  \int\frac{\dif q^+}{4\pi}\int\dif k^+ \; \frac{q^+}{k^+}
  \esp{-\ui\frac{q^+}{2}x^-} \,\Theta(x^+) \;.
\end{equation}

The $q^+$-dependent part, by dividing out the rapidity factor
$(2/\pi)\int\dif k^+/k^+ = 2Y/\pi$ (as requested in order to obtain the real
part of the diagram from the dispersion relations) and the leading field
$a_0(r^2)$ fits with the displacement form
$\ui(\Delta_R+\ui\Delta_I)\ui\partial_- = -(\Delta_R+\ui\Delta_I)\partial_-$,
by thus providing the shift $2\pi R \Delta_H$ from past to future, where
\begin{equation}\label{DeltaHb}
  \Delta_R = \Delta_H = -\frac1{a_0(r^2)} \frac{(\pi R)^2}{2}\int\dif^2\xt'\;|h_0(\xt')|^2
  \frac{\dif^2\qt}{2\pi^2\qt^2}\;\esp{\ui\qt\cdot(\xt-\xt')} \;.
\end{equation}
We recognize here the 2-dimensional Laplacian Green function $G_0(\xt-\xt')$
applied to the H-diagram density. By translating it to the azimuthal-averaged
formalism we can use the one-dimensional form
\begin{equation}\label{lgf}
  G_0(r,r') = \frac1{2\pi}\log\frac{L^2}{r_>^2} \;, \qquad
  \big(r_> \equiv \max(r,r')\big)
\end{equation}
and apply it to the rescattering current $\Jr(r^2)$. The result is just
identical to the r.h.s.\ of eq.~(\ref{solDelta}) with the $\Delta_0$
contribution subtracted out. In particular, for $r\gg b$ eq.~(\ref{DeltaHb})
factorizes and yields $\Delta_H(\infty)=-a_H$, thus recovering the
form~(\ref{Akin}) of the total shift.

Our overall interpretation of the present findings is that a large-distance
observer will see the particle shock-waves to suffer a total time delay
$2\pi R\,\As(b)$, directly related to the action and thus to the scattering
angle
\begin{equation}\label{sinth}
  \sin\frac{\theta_\cl(b)}{2} = -\pi R\frac{\partial}{\partial b} \As(b)
  = -\frac1{\sqrt{s}}\frac{\partial}{\partial b} \A_R = \frac{bR}{\rho(b^2)}
\end{equation}
because of the contribution of the rescattering shift. On the other hand, a test
particle parallel to particle 2, traveling at transverse distance $r\simeq b$
from particle 1 
\footnote{We mean $r=b$ at deflection time, corresponding to the perpendicular
  distance (impact parameter) $b_0=b\cos(\theta/2)$, according to the
  discussion of sec.~\ref{s:wpm}.}
will see a profile function $2\pi R\,a(r^2)$ related to its own scattering angle
\begin{equation}\label{sintest}
  \sin\frac{\theta(r)}{2} = -\pi R \frac{\partial}{\partial r} a(r^2) =
  \frac{rR}{\rho(r^2)} \xrightarrow{r\to b} \sin\frac{\theta_\cl}{2}
\end{equation}
which turns out to be the same in the limit $r=b$.%
\footnote{That is because the $b$-derivative of $\A_R(b)$ has implicit
  contributions which vanish by the equation of motion.}
Furthermore, the test-particle will suffer a time delay that, due to the
double-shift picture of sec.~\ref{s:sdts}, is $\Delta t-\Delta z=2\pi R\,a(x)$,
close to the short-distance value $2\pi R\,a(b)=\partial\A_R/\partial\sqrt{s}$
coming from the eikonal insertions of sec.~\ref{s:ssw}

Finally, while the rescattering contribution is needed for the large distance
shifts, its contribution to the action starts at order $(R/b)^4$ because the
diagram in fig.~\ref{f:sublDiagrams}.c (formally of order $(R/b)^3$) vanishes by
the property $\sgn(x^+)\delta(x^+)=0$. Therefore it does not affect the
consistency of the metric and action descriptions at order $(R/b)^2$.%
\footnote{The absence of corrections of relative order $(R/b)^3$ was
  argued~\cite{ACV88,ACV93} to vanish for light-particle scattering by
  analiticity and relativity arguments.}

Of course, a symmetrical calculation yields the shift
$\pi R\sgn(x^-)\bar\Delta(\xt)$ for the field $h_{++}$, provided one exchanges
light-cone indices $+\leftrightarrow -$ and the profile functions and shifts
$a\leftrightarrow \bar{a}$, $\Delta\leftrightarrow \bar{\Delta}$.

%%%%%%%%%%%%%%%%%%%%%%%%%%%%%%%%%%%%%%%%%%%%%%%%%%%%%%%%%%%%%%%%%%%%%%%%%%%%%%%%%
\subsection{Shift modification for the transverse field\label{s:smtf}}
%%%%%%%%%%%%%%%%%%%%%%%%%%%%%%%%%%%%%%%%%%%%%%%%%%%%%%%%%%%%%%%%%%%%%%%%%%%%%%%%%

We now go over to the second basic equation~(\ref{eqphi}) which essentially
describes rescattering properties (fig.~\ref{f:sublDiagrams}.c,d) of the
transverse field $h=\nabla^2\phi$, once emitted by the longitudinal fields. We
note that, by replacing the RAM fields~(\ref{shifts}) in the r.h.s.\ of
eq.~(\ref{eqphi}), we obtain the structure
\begin{align}
  \kappa(\kappa\sqrt{s})^2 &\big[\dsq(\bar{a}\dq\phi)+\text{c.c.}\big]
  \times  \label{eqphiStr} \\ \nonumber
  &\left\{
  -\frac{\partial}{\partial\Delta}\frac12\Big[
   \delta(x^- -\pi R\Delta)\delta(x^+ -\pi R\bar\Delta_\phi)
  +\delta(x^- +\pi R\Delta)\delta(x^+ + \pi R\bar\Delta_\phi)\Big]\right\} \;.
\end{align}

Let us recall that, according to eqs.~(\ref{Delta0},\ref{solDelta}), the
longitudinal rescattering predicts a shift $\Delta=a(b)$. We then decide to
expand the shift $\Delta_\phi$ of the transverse field around the constant value
$\Delta=a(b)\neq0$, which gives the leading order of the longitudinal shift. By
this method, it is easy to evaluate a (formally) first order modification
$\Delta^{(1)}\equiv\Delta-\Delta_\phi$ of the transverse-field shift, which
however is also of leading order in the $R/b$ expansion.  By expanding the first
term in the l.h.s.\ of eq.~(\ref{eqphi}) we obtain
\begin{align}
  \partial_-\partial_+ |\partial|^4 \Phi
 &= |\partial|^4 \phi\left[ 1 -\Delta^{(1)} \frac{\partial}{\partial\Delta}
   -\bar\Delta^{(1)} \frac{\partial}{\partial\bar\Delta}+\cdots \right]
 \times \nonumber\\
& \frac12\big[
  \delta(x^- -\pi R\Delta) \delta(x^+ -\pi R\bar\Delta)
  +\delta(x^- +\pi R\Delta)\delta(x^+ + \pi R\bar\Delta) \big]\label{expPhi}
\end{align}
so that we can easily match the left and the r.h.s.\ to get the equation, in the
axisymmetric limit,
\begin{equation}\label{eqDelta1}
  2|\partial|^4 (\phi\Delta^{(1)}) = \dsq(\bar{a}\dq\phi) + \text{c.c.} \;,
\end{equation}
where we work in the regime $r \gg b$ with
$\bar{a}(r^2) = a(r^2) \simeq\frac1{2\pi}\log\frac{L^2}{r^2}$.

The expression of $\phi$ is found by the RAM equation
\begin{equation}\label{eqzphi}
  -2\frac{\dif}{\dif r^2} \frac{r^2\ddot\rho}{(2\pi R)^2} = 2|\partial|^4\phi
  = \dsq a\,\dq\bar{a} - \dmq a \,\dmq\bar{a} = -\frac{\dif}{\dif r^2}
  (r^2\dot{a}\dot{\bar{a}}) \;,
\end{equation}
which was provided before~(\ref{eomrho}). By inserting proper boundary
conditions, in our regime $r\gg b$ we can set $\phi=\frac14[a(r^2)-a(b^2)]^2$
and tentatively look for a solution $\Delta^{(1)}=\lambda a(r^2)+\mu$, where
$\lambda$ and $\mu$ are constants. A simple calculation shows that
\begin{equation}\label{simpleDelta}
  \Delta^{(1)} = \frac23 a(r^2) + \frac13 a(b^2)-\frac1{2\pi} \;, \qquad
  \Delta_\phi = \frac23 [a(b^2)-a(r^2)] + \frac1{2\pi} \;,
\end{equation}
so that $\Delta_\phi$ is $r^2$-dependent and of leading order, while the
correction $\Delta^{(1)}$ is sizeable. Furthermore, $\Delta_\phi$ becomes cutoff
independent, and pretty small at $r=b$.

The above conclusion may be unpalatable from a calculational standpoint, but is
natural on physical grounds because rescattering occurs for the $\phi$ field at
relatively leading level and at all distances, so that the parameter $a(b)$ has
no particular role and a sizeable difference $\Delta-\Delta_\phi$ is
expected. One might ask, at this point, whether the RAM hierarchy for the action
is really satisfied or not. Fortunately it remains, because the integral over
$x^+$, $x^-$ of the rescattering vertex with the longitudinal field just
vanishes ($\sgn(x^-)\delta(x^-)=0$) or, in other words, there is no $R^3/b^3$
contribution to the action from diagram~\ref{f:sublDiagrams}.c. Furthermore, the
shift does not change the $x^+,\;x^-$ integration in the H-diagram and therefore
rescattering contributions.

On the other hand, the location of the shock-wave and the evolution of particle
trajectories and geodesics does change from the point of view of the metric, so
that rescattering is needed in order to have self-consistent calculations even
at relative order $R^2/b^2$.

%%%%%%%%%%%%%%%%%%%%%%%%%%%%%%%%%%%%%%%%%%%%%%%%%%%%%%%%%%%%%%%%%%%%%%%%%%%%%%%%%
\section{Irregular solutions and their (re)scattering\\ properties\label{s:isrp}}
%%%%%%%%%%%%%%%%%%%%%%%%%%%%%%%%%%%%%%%%%%%%%%%%%%%%%%%%%%%%%%%%%%%%%%%%%%%%%%%%%

All preceding arguments hold for the UV-safe solutions of the RAM model in
eqs.~(\ref{eoma},\ref{eomrho}) which, by definition satisfy the condition
$\rho(0)=0$. Due to the form of $\rho(r^2)$ in eq.~(\ref{rho}), $\rho(0)\neq0$
would imply that $\dot\phi\sim-\rho(0)/r^2$ has a short-distance singularity,
$\gr\phi$ has an outgoing flux $-\rho(0)$ and therefore
$h=\gr^2\phi\sim-\rho(0)\delta(r^2)/(\pi R)^2$ has a singular $\delta$-function
contribution. The latter behaviour would call for large short-distance effects
--- possibly regularized by the string --- which are expected, but not
considered, in the RAM model.

Restricting the solutions by $\rho(0)=0$ is possible only if the impact
parameter is larger than some critical value $b_c\sim R$ which signals a
possible classical collapse. In fact, the regular solution~(\ref{regsol}) of
eq.~(\ref{eomrho}) has continuous derivative at $\tau\equiv r^2=b^2$ provided
\begin{equation}\label{tbcond}
  t_b(1-t_b^2) = \frac{R^2}{b^2} \;,
\end{equation}
a condition which has real-valued solutions only for
$b^2\geq b_c^2=(3\sqrt{3}/2)R^2$. On the other hand, the complex solutions for
$b<b_c$ lead to an exponential damping~\cite{CC08} of the $S$-matrix of type
$\sim\exp(-Gs)$ --- with exponent of the order of a black-hole entropy $ER$ ---
which leads, eventually, to a violation of unitarity of the RAM
model~\cite{CC09,CiCoFa}.

From the above discussion, some questions arise: What happens to the
rescattering solutions for $b<b_c$, when they are irregular? Do we see any sign
of a possible collapse in the latter, perhaps in relation with the unitarity
problem and to a ``fall in the center'' mechanism~\cite{DDRV10}?

Let us then consider a class of real-valued RAM solutions with $\rho(0)\neq0$ for
$b<b_c$, as follows. We start from the general solution for $\rho(0)\neq0$
\begin{equation}\label{irrsol}
  \rho(\tau) = [\rho(0)+t_b \tau] \Theta(b^2-\tau)+R^2\cosh^2\chi(\tau)\Theta(\tau-b^2)
\end{equation}
where $t_b$ is now a free parameter which $\rho(0)>0$ depends on. Then we
choose $t_b\equiv t_m$ so as to minimize $\rho(0)$, i.e.,
$\dif\rho(0)/\dif t_b|_{t_m}=0$. The parameters of this ``minimal'' solution
$\rho_m(\tau)$ are given by
\begin{equation}\label{minpar}
  \frac{(1-t_m^2)^2}{2t_m} = \frac{R^2}{b^2} \;, \qquad
  \rho_m(0) = \frac{R^2}{1-t_m^2}-t_m b^2 
  = R^2\frac{1-3t_m^2}{(1-t_m^2)^2} \;.
\end{equation}
The corresponding rescattering current is
\begin{equation}\label{irrJR}
  \Jr(\tau) = (2\pi R)^2(\tau\ddot\phi)^2 = \frac1{(2\pi R)^2}\left(\dot\rho-
    \frac{\rho}{\tau}\right)^2  \xrightarrow{\tau<b^2}
  \frac1{(2\pi R)^2}\frac{\rho^2(0)}{\tau^2} \;.
\end{equation}
The above current is now non-vanishing for $\tau<b^2$, and, in addition, has a
non-integrable behaviour for $\tau\to0^+$.
This means that large rescattering amplitudes are built in the short-distance
region $\lambda_s < r,b < R$ and, as a consequence, that exchange and emission
of (massive) string states can no longer be neglected.

Nevertheless, we find it instructive to provide here a preliminary analysis of
irregular solutions in our effective-theory framework in which only graviton
intermediate states are considered. Basic string effects –-- like graviton
reggeization and ensuing string production at $\lambda_s$~\cite{ACV88}, as well
as diffractive and central string emission induced by tidal
forces~\cite{ACV88,GiGrMa08}
\footnote{Tidal-force production amplitudes are determined by the second
  $(b,r)$-derivatives of phaseshifts~\cite{ACV88,GiGrMa08} and are thus possibly
  important in the whole subcritical region $\lambda_s < r,b < R$ where
  (\ref{irrJR}) is large, so that unitarity defect~\cite{CC09,CiCoFa} and energy
  balance~\cite{Rychkov_pc,CCGV} could be sizeably affected.}
should certainly be estimated in the near future in order to see how they affect
the picture, but we feel that our effective approach may still provide
suggestions and questions to be answered, and is anyway needed as a ground for
the estimates just mentioned.

In order to get a better insight on the evolution of irregular solutions, we cut-off
the current (\ref{irrJR}) below $r=\lambda_s\ll R$, and we compute
the total charge, related in sec.~\ref{s:rcram} to the H-diagram action $2\pi
Gs\,a_H$ and to the shift $\Delta_H$ as follows:
\begin{align}
  a_H &= -\Delta_H = 2\pi\int_{\lambda_s^2}^\infty \Jr(\tau)\;\dif\tau \nonumber\\
  &= \frac1{2\pi R^2}\left[\int_{0^+}^\infty (1-\dot\rho)^2\;\dif\tau
  -2[1-\dot\rho(0^+)]\rho(0) + \frac{\rho^2(0)}{\lambda_s^2}
  +\ord{\lambda_s^2}\right] \;, \label{aHirr}
\end{align}
where the singular integration has been performed by a careful integration by
parts and in the last equality we have introduced the $\delta$-function
singularity in the $\rho$-derivative:
\begin{equation}\label{irrdrho}
  \dot\rho(\tau) = \dot\rho(\tau)|_\reg +\rho_m(0)\delta(\tau) \;, \qquad
  \gr^2\phi=\frac{1-\dot\rho(\tau)|_\reg}{(\pi R)^2} -
  \frac{\rho(0)\delta(\tau)}{(\pi R)^2} \;.
\end{equation}
This means that we think of $\rho(\tau)=0$ for $\tau\leq 0$ (a region which is
felt by the two-dimensional model through the outgoing flux) and that the
$\delta$-function occurs because of the $\tau$-discontinuity.

Thus, strictly speaking, we discover that the singular solutions are not {\em
  bona fide} solutions of the equations of motion up to $\tau=0^-$ unless we
introduce a singular ``external force'' $\ddot\rho=\rho(0)\delta'(\tau)$ at
$\tau=0$. We can do that by starting from the action~\cite{CiCoFa}
\begin{equation}\label{2dimAction}
  \frac{\A}{2\pi Gs} = \int\dif^2\xt\left[ a\bar{s}+\bar{a}s
  -\frac12\gr a\cdot\gr\bar{a}-(\pi R)^2 \left(\frac12(\gr^2\phi)^2
   +\gr\phi\cdot\gr\Hc\right)\right]
\end{equation}
and by supplementing the Lipatov's current $\Hc$ of eq.~(\ref{defHc}) with a
singular external current
\begin{equation}\label{deltaHc}
  \delta\dot\Hc = \dot\Hc+2\dot{a}\dot{\bar{a}} = -\frac{\rho_m(0)}{(\pi R)^2}
  \delta'(\tau) \;.
\end{equation}
By then replacing such expression into eq.~(\ref{2dimAction}), i.e.,
\begin{equation}\label{grgr}
  -(\pi R)^2\gr\phi\cdot\gr\Hc = -[\rho(\tau)-\tau]\left(
    2\dot{a}\dot{\bar{a}}+\frac{\rho_m(0)\delta'(\tau)}{(\pi R)^2}\right) \;,
\end{equation}
we find that the action enforces eq.~(\ref{irrdrho}) and becomes, on the equation
of motion,
\begin{align}
  \frac{\A}{2\pi Gs} &= \pi\left\{\int\left[a\bar{s}+\bar{a}s
      -2\rho\dot{a}\dot{\bar{a}}\right]\dif\tau-\frac12\int_{0^+}^\infty
    \left(\frac{1-\dot\rho}{\pi R}\right)^2\dif\tau
    +\frac12\left(\frac{\rho_m(0)}{\pi R}\right)^2\delta(0) \right\} \nonumber\\
  &= a(b) - \int_{0^+}^\infty \frac{(1-\dot\rho)^2}{2\pi R^2}\;\dif\tau
  +\frac{\rho_m^2(0)}{2\pi R^2\lambda_s^2} \;, \label{irrAct}
\end{align}
thus determining the singular term in the action.%
\footnote{Note that we interpret the distribution $\delta(\tau)$ as a
  step-function located around $\tau=0$ with width $\lambda_s^2$ and height
  $\lambda_s^{-2}$, hence the identification $\delta(0)=\lambda_s^{-2}$.}

We are now able to look at the scattering and shift properties of the irregular
solutions, by keeping in mind that we need in this case the external
current~(\ref{deltaHc}), that we think generated by the short-distance string
dynamics and/or by possibly collapsed matter. Compared to the complex solutions
--- which have a quantum-tunneling interpretation and cause a probability
suppression --- the irregular ones may provide a probability source or
alternatively may carry away the information loss.

\begin{figure}[t!]
  \centering
  % graphics from ~/work/gravitunneling/irregularAction.gp
  \includegraphics[height=0.5\linewidth,angle=270]{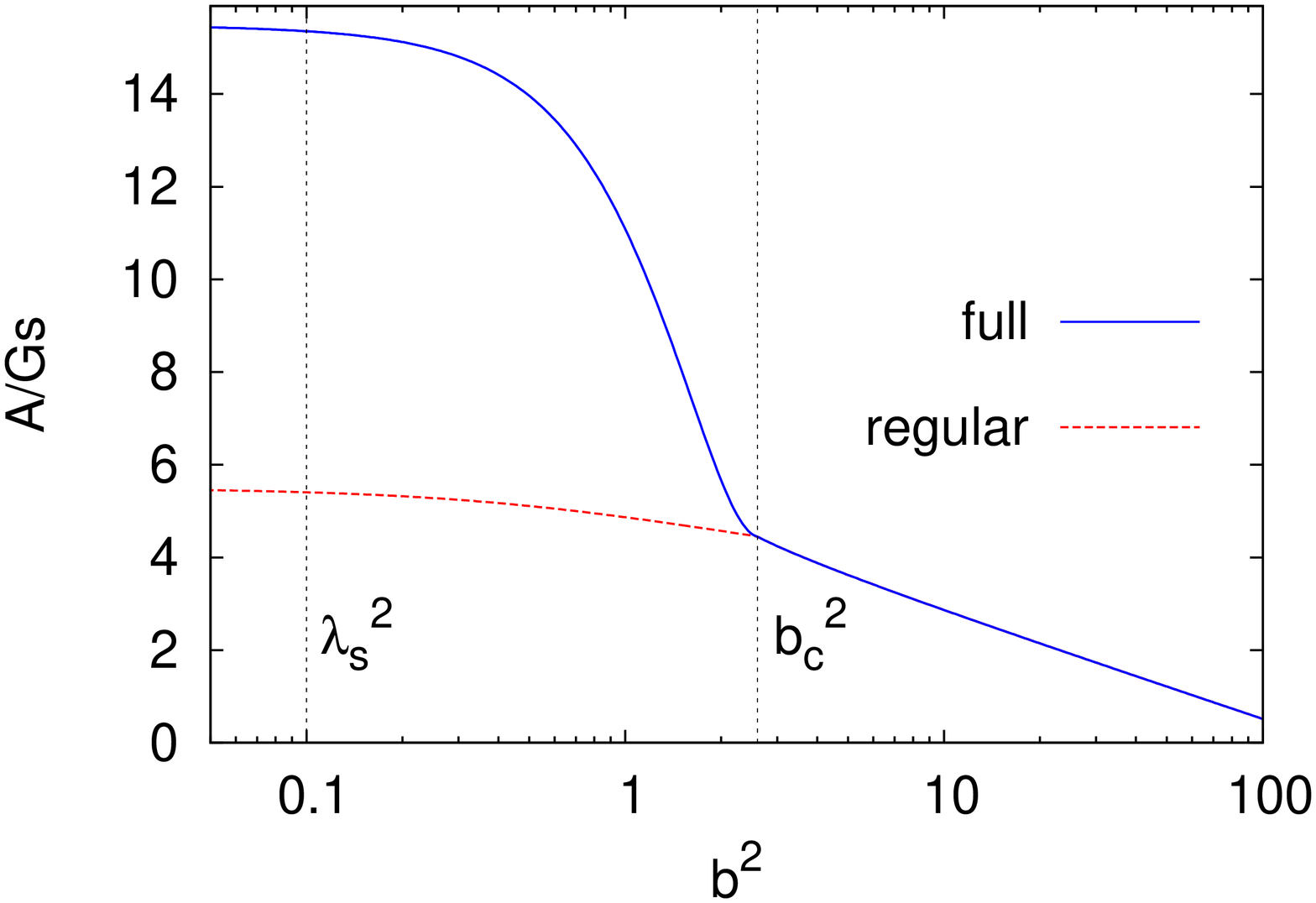}%
  \includegraphics[height=0.5\linewidth,angle=270]{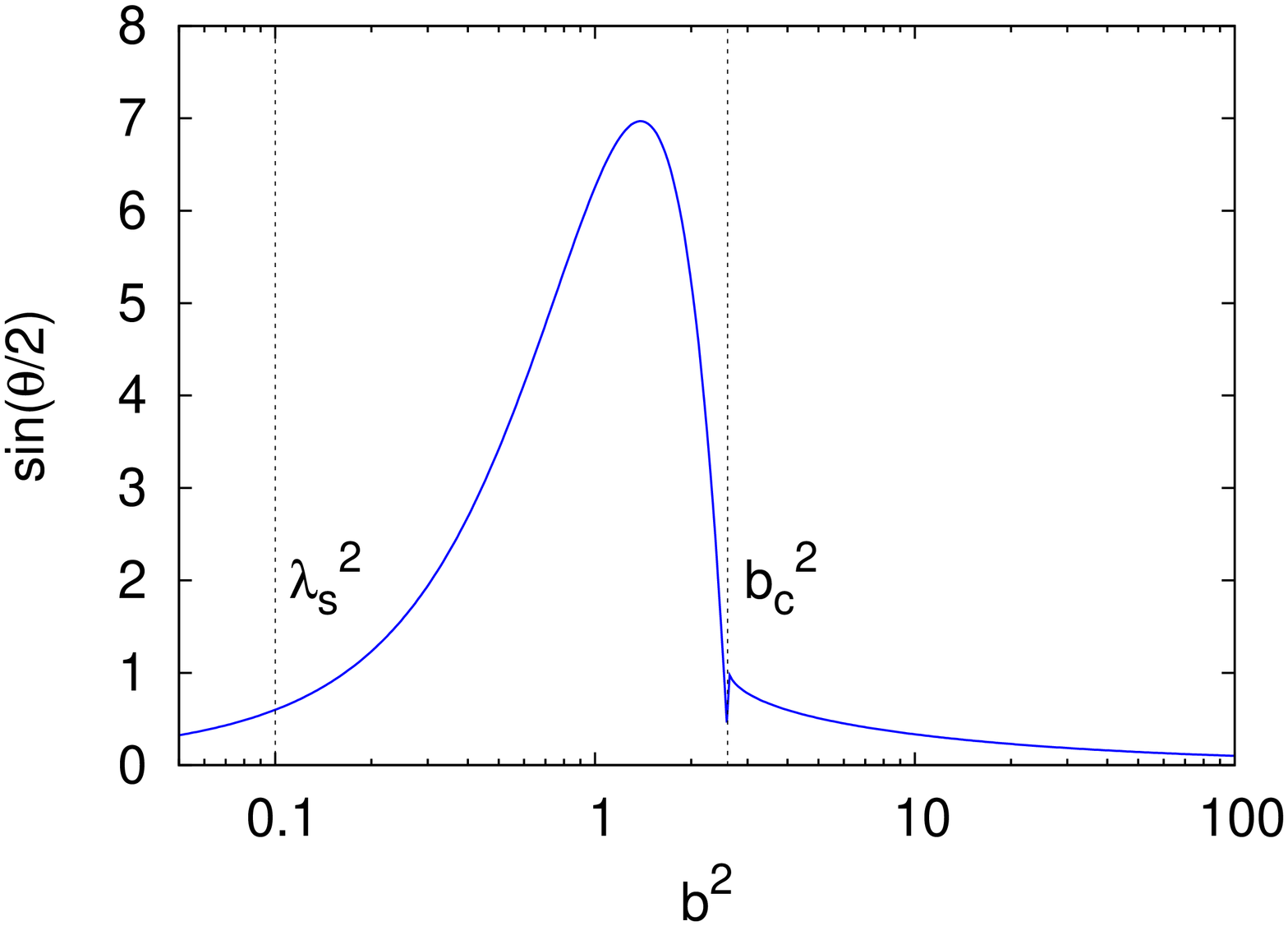}
  \Caption{{\em (a)} action of the solutions of equation of motion (blue solid line);
    the red dashed line shows the action for $b<b_c$ without the $\rho(0)$
    contribution; {\em (b)} scattering angle as derived by eq.~(\ref{irrTheta}).  In
    both plots $\lambda_s^2 = 0.1 R^2$; lengths are measured in units of $R=1$.}
  \label{f:irrAction}
\end{figure}

By thus using the eikonal representation of the $S$-matrix with the
action~(\ref{irrAct}) and by carefully computing $b$- and
$\sqrt{s}$-derivatives, we get the scattering angle
\begin{equation}\label{irrTheta}
  \pm\sin\frac{\theta_s}{2} = -\frac1{\sqrt{s}}\frac{\partial\A}{\partial b}
  = \frac{Rb}{\rho(b^2)} + \pi^2 R \Hc(0) \frac{\partial\rho(0)}{\partial b}
\end{equation}
and the time delay
\begin{equation}\label{irrDelay}
  t_D = \frac{\partial\A}{\partial\sqrt{s}} = 2\pi R \left[ a(b)
    -\frac{\pi}{2}\Hc(0)\sqrt{s}\frac{\partial\rho(0)}{\partial\sqrt{s}}\right]
  \equiv 2\pi R\Delta_0 \;,
\end{equation}
where, by eqs.~(\ref{irrsol},\ref{deltaHc})
\begin{equation}\label{irrH0}
  \Hc(0) = \frac1{(\pi R)^2}\left[1-\dot\rho(0^+)-\frac{\rho(0)}{\lambda_s^2}
  \right] \;.
\end{equation}
Such results can be further specified by using
\begin{equation}\label{spDer}
  \frac{\partial\rho(0)}{\partial b} = -2 t_m b =
  -\left(\frac{Rb}{\rho(b^2)}\right)^2 b \;, \qquad
  \frac{\sqrt{s}}{2}\frac{\partial\rho(0)}{\partial\sqrt{s}} = \frac{R^2}{1-t_m^2}
  = \rho(b^2)
\end{equation}
and the outcomes are plotted in fig.~\ref{f:irrAction}.

We note the strong increase of the action in the region $b\simeq\lambda_s$ due
to the positive singular contribution --- implying in particular a motion with
many turns in the region $\lambda_s \lesssim b \lesssim R$. Actually, large
scattering angles are reached pretty soon around $b=b_c$ as solutions of the
equation
\begin{equation}\label{deflec}
  \pm\sin\frac{\theta_s}{2} = \left(\frac{b}{R}\right)^3 \frac{R^2}{\lambda_s^2}
  \left[1-3 t_m^2\Big(\frac{b}{R}\Big)\right] \;,
\end{equation}
but, due to the strong increase of the action derivative (fig.~\ref{f:irrAction}), the
saddle-points are confined to either the critical region
$b_c^2-b^2=\ord{\lambda_s^2}$ or to the small-$b$ region
$(b/R)^3=\ord{\lambda_s^2/R^2}$, for all real values of $\theta_s$.%
\footnote{We have checked that complex saddle-points do not change the picture.}
Finite $b$-values, in the region $\lambda_s^2 \ll b^2 < b_c^2-\lambda_s^2$ are
strongly suppressed like $\exp(-Gs R^2/\lambda_s^2)$.

The above observation suggests that the singular solutions under study do not
actually yield back all the initial information in the physical scattering
region, but may carry it away in the small-$b$, string-dominated region.

Another piece of information comes from the study of the shifts and the
corresponding time delay. We first notice, by eq.~(\ref{irrDelay}), that the
short-distance shift of
\begin{equation}\label{irrDelta0}
  \Delta_0 = a(b) +
  \left[\frac{\rho_m(0)}{\lambda_s^2}-(1-t_m)\right]\frac{2\rho(b^2)}{2\pi R^2}
\end{equation}
has a positive singular term --- meaning a lot of time spent in the interaction
region --- which is essentially twice the singular contribution to $a_H$ in
eq.~(\ref{aHirr}). The factor of 2 comes from $\rho(0)$ scaling like
$R^2\sim (G\sqrt{s})^2$ --- similar to what happens for the H-diagram itself.

Furthermore, we already know from sec.~\ref{s:rcram} that the large-distance
shift takes contributions from rescattering insertions and produces the
additional (negative) shift $\Delta_H$, the one we started with in
eq.~(\ref{DeltaH}). We thus have
\begin{align}
  \Delta(\infty) &= \Delta_0 + \Delta_H = a(b) - \int_{0^+}^\infty 
  \frac{(1-\dot\rho)^2}{2\pi R^2} \; \dif\tau +
  \frac1{2\pi R^2}\frac{\rho_m^2(0)}{\lambda_s^2} - \Hc(0)\pi t_m b^2
  \nonumber\\
 &= \As(b;\lambda_s) - \Hc(0)\pi t_m b^2 \;.
\end{align}
We note that the singular term remains positive after subtraction of $\Delta_H$
and of the same order as that of the action --- apart for the addition of a term
proportional to $-\Hc(0) > 0$. The very large time delay is thus confirmed at
all distances. That means that the singular solutions, wherever they are, spend
a long time in the interaction region, before exiting, either around $b=b_c$, or
at $b=\ord{\lambda_s}$. Therefore, they are connected with long-lived states in
that region.

Finally, let us look at the associated metric~(\ref{metr}), whose geodesics
--- in the ``regular'' case $\rho(0)=0$ --- were argued to provide a
complementary picture of particle scattering. For the irregular solutions of
eqs.~(\ref{irrsol},\ref{minpar}), the relevant metric coefficients (inside the
light-wedges $x^+ x^- > 0$) have the form
\begin{subequations}\label{Gmunu}
  \begin{align}
    G_{+-} &\equiv -2 g_{+-} = 1 -\frac12 (1-\dot\rho) &\longrightarrow&\quad 1 \\
    G_{rr} &\equiv g_{rr} = 1 +\frac12 (1-\dot\rho) -\frac12
    \left(\frac{\rho}{r^2}-\dot\rho\right) &\longrightarrow&\quad
    1-\frac{r^*{}^2}{r^2}  \qquad \left(r^*{}^2 \simeq \frac{\rho(0)}{2}\right)\\
    G_{\theta\theta} &\equiv\frac{g_{\theta\theta}}{r^2} = 1 +\frac12 (1-\dot\rho)
    +\frac12 \left(\frac{\rho}{r^2}-\dot\rho\right) &\longrightarrow& \quad
    1+\frac{r^*{}^2}{r^2} \;,
  \end{align}
\end{subequations}
where the arrows label the simplified expressions obtained in the $\dot\rho\to1$
limit, which will be used for the purpose of the qualitative discussion below.

We note that, for $\rho(0)\neq0$, $g_{rr}$ shows a puzzling zero at
$r^2 = r^*{}^2 \sim \rho(0) \sim R^2$, which causes a change of signature of the
$r$-dimension for $r<r^*$ --- a feature to be taken with great caution because
our perturbative identification of the metric coefficient is probably invalid in
the strong-coupling region we are interested in. Nevertheless, let us take the
expressions~(\ref{Gmunu}) at face-value, and discuss the ensuing geodetic flow,
in the transverse plane, which is strongly affected by the $\phi$-field
singularity $\dot\phi \sim -\rho(0)/r^2$.

In order to do that, we shall introduce the test-particle lagrangian (per unit mass)
\begin{equation}\label{tplag}
  L = -\frac{|\dif s|}{\dif t} = -\sqrt{G_{+-}(1-\dot{z}^2)-G_{rr}\dot{r}^2
  - G_{\theta\theta} r^2 \dot{\theta}^2}
\end{equation}
with corresponding momenta and hamiltonian (per unit mass)
\begin{subequations}\label{momenta}
  \begin{align}
  P_z &\equiv\frac{\dif L}{\dif\dot{z}} = \frac{G_{+-}}{-L}\dot{z} =
  \text{const} \\
  P_\theta &\equiv\frac{\dif L}{\dif\dot{\theta}} =
  \frac{G_{\theta\theta}}{-L} r^2\dot{\theta} = \text{const} \\
  P_r &\equiv\frac{\dif L}{\dif\dot{r}} = \frac{G_{rr}}{-L}\dot{r}\\
  H &\equiv P_z \dot{z} + P_\theta \dot\theta + P_r \dot{r} - L
  = \frac{G_{+-}}{-L} = \text{const} \;.
  \end{align}
\end{subequations}
It is convenient to define the momenta per unit energy $p_i \equiv P_i/H$, so as
to allow the treatment of the massless case too. By evaluating the constants of
motion in the asymptotic region $r\to\infty$ one finds that $p_z=\dot{z}=v_z$,
$H = (1-v^2)^{-1/2}$ and $p_\theta = \beta v_\perp$, where $v_z$, $v_\perp$ and
$v=\sqrt{v_z^2+v_\perp^2}$ are the longitudinal, transverse and total (asymptotic)
velocity, while $\beta$ is the impact parameter of the test particle w.r.t.\ the
$z$ axis.

The test-particle motion is eventually described by the constants of motion
$p_z$, $p_\theta$, $H$ and by the radial ``effective potential''
\begin{align}\label{potential}
  -p_r^2 = -\frac{G_{rr}}{G_{+-}}\left[1-v_z^2-\left(
    1-v^2+\frac{(\beta v_\perp)^2}{G_{\theta\theta}r^2}\right)G_{+-}\right]
  \longrightarrow -\left(1-\frac{r^*{}^2}{r^2}\right) v_\perp^2\left(1
   -\frac{\beta^2}{r^2+r^*{}^2}\right) \;,
\end{align}
since $\dot{r}^2 -p_r^2 (G_{+-}/G_{rr})^2 = 0$.

\begin{figure}[t!]
  \centering
  % graphics from ~/Matematica/gravitunneling/metricaSingolare.nb
  \includegraphics[width=0.57\linewidth]{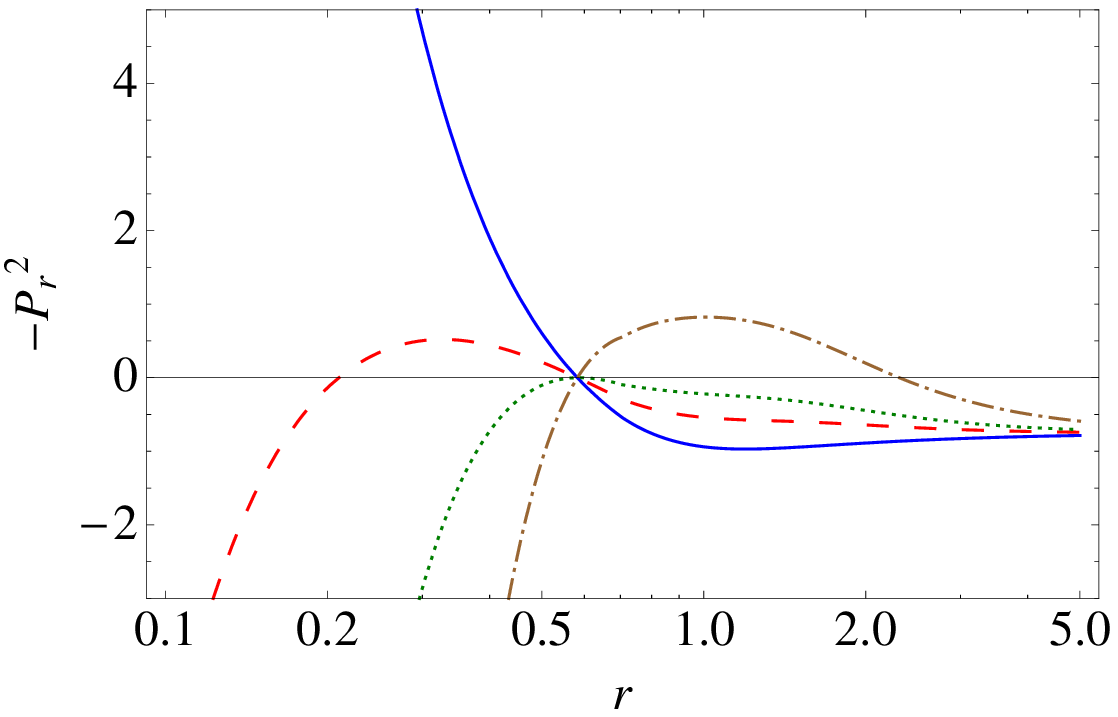}
  \Caption{Radial ``effective potential'' for different values of the
    test-particle's impact parameter: $|\beta|<r^*$ (solid blue),
    $r^*<|\beta|<\sqrt{2}r^*$ (dashed red), $|\beta|=\sqrt{2}r^*$ (dotted
    green), $|\beta|>\sqrt{2}r^*$ (dash-dotted brown).}
  \label{f:potential}
\end{figure}
The $\beta=0$ geodesics are characterized by the fact that the transverse
velocity $v_\perp$ (if present) is purely radial and the
potential~(\ref{potential}) is purely repulsive (fig.~\ref{f:potential}), and
would be typical of an angular momentum barrier with
$|p_\theta/v_\perp|=|\beta|=r^*$ that --- by lack of other explanations --- we
should attribute to the external source $\delta\Hc$ in eq.~(\ref{deltaHc}) that we
have put in at $r=0$ in order to justify the irregular solutions themselves. For
$v_\perp=0$ the repulsive potential is absent, and the geodesics (at $x^-=0$,
say) will reach the shock-wave (at $x^+=0$, say) at the initial distance $r$,
but will thereby acquire a shift and will be deflected according to the formula
\begin{equation}\label{refraction}
  \tan\frac{\theta_f}{2} = \frac1{\sqrt{1-\frac{r^*{}^2}{r^2}}} \left(
  \tan\frac{\theta_i}{2} - \pi R \frac{\dif\bar{a}(r)}{\dif r}\right) \;,
\end{equation}
which takes into account the ``refraction index'' of the $\phi$-field. Once
again $r=r^*$ is unreachable: if initially we try to approach it, the geodesic
scattering angle approaches $\pi$, and the test particle is reflected backwards.

We note at this point that the situation does change --- unexpectedly --- when
$|\beta|$ increases and overcomes the threshold $r^*$: in this case, according
to eq.~(\ref{potential}), the centrifugal factor at the origin changes sign, so
that the potential becomes partly attractive, close to the origin and above
$r^*$ (fig.~\ref{f:potential}). In particular, for $|\beta|=\sqrt{2}r^*$ the
potential is everywhere attractive!

This somewhat surprising feature opens up the possibility that test particles
may reach the small-$r$ region, at least at quantum level, because the small
barrier has a transmission coefficient of order unity around the value
$|\beta|=\sqrt{2}r^*$, and is thus able to populate the small-$r$ region around
$\lambda_s$ in an efficient way. That feature is the counterpart of the
selection of possible issues of the particles themselves, which was argued
before to have sizeable probability for a small-$b$ exit on the basis of the
action stationarity condition. Furthermore, the test-particle action variable
$\frac{1}{p_\perp}\int_{\lambda_s}^{r^*} p_r\;\dif r\sim\sqrt{2}r^*\log\frac{r^*}{\lambda_s}$
develops a logarithmic behaviour in the small-$r$ region with frequency
$(\sqrt{2}r^*)^{-1}\simeq R^{-1}$: is that perhaps a possible interpretation of
the time nature of the variable $r$ in that region?

On the whole, we think, the above discussion shows --- at a sort of
``phenomenological'' level --- that the irregular solutions may indeed vehicle
the particles' information from distances of order $R$ down to the string size,
thus suggesting that the unitarity loss of the $\rho(0)=0$ model is due to the
opening up of the gate to other worlds. However, due to the lack of the
theoretical ingredients from string theory mentioned in the beginning, in the
present formulation we have been unable to really assess the reliability of the
above suggestion and to discuss the physical interpretation of the state(s),
living at $\lambda_s$, which could be responsible for the singular contributions
to the action investigated here and for their intriguing consequences.

%%%%%%%%%%%%%%%%%%%%%%%%%%%%%%%%%%%%%%%%%%%%%%%%%%%%%%%%%%%%%%%%%%%%%%%%%%%%%%%%%
\section{Conclusions\label{s:conc}}
%%%%%%%%%%%%%%%%%%%%%%%%%%%%%%%%%%%%%%%%%%%%%%%%%%%%%%%%%%%%%%%%%%%%%%%%%%%%%%%%%

To sum up, in our study of the improved eikonal model with subleading correction
we have found that:
\begin{itemize}
\item The source particles, after the interaction, besides being deflected,
  suffer a time delay which can be interpreted as a shift in the light-cone
  variables~(sec.~\ref{s:ilepp});
\item The shock-wave metric fields $h_{\pm\pm}$ generated by the source
  particles are also shifted in an analogous way, after the interaction described
  by the eikonal amplitude~(sec.~\ref{s:ssw});
\item The subleading contributions to the eikonal, represented by the (multi)
  H-diagram and by the rescattering terms, provide corrections to the scattering
  angle of the particles and to the shifts of the metric fields $h_{\pm\pm}$ and
  $\Phi$~(sec.\ref{s:rcram}). Due to rescattering, the large-distance shift of
  the particle fields agrees with the scattering-angle description based on the
  action.
\end{itemize}
In order to find an adequate form for the self-consistent metric which embodies
all these features, we have to modify the ACV expression~(\ref{metr}).

As far as the shock waves are concerned, the improved expressions are obtained
from eq.~(\ref{hppR}) by replacing the leading profile function $a_0(\xt)$ with
the full profile function $a(\xt)$ derived in sec.~\ref{s:Hd} and by
accordingly shifting the location of the shock waves of the amount
$\pm\pi R\Delta(\xt)$ as explained in sec.~\ref{s:rssf}.

In turn, the support $\sphi$ of the transverse field $\Phi$ is also shifted from the
region $x^+ x^- > 0$ to the two (past and future) disconnected regions
\begin{align}
  \sphi^{(-)} &\equiv \{x\;:\; x^+ < -\pi R \bar\Delta_\phi(\xt) \;,\;
   x^- < -\pi R \Delta_\phi(\xt)\} \nonumber\\
  \sphi^{(+)} &\equiv \{x\;:\; x_R^+ > +\pi R \bar\Delta_\phi(\xt_R) \;,\;
   x_R^- > +\pi R \Delta_\phi(\xt_R)\}\label{sphi}
\end{align}
where in the region $\sphi^{(+)}$ after the interaction we have used the rotated
coordinates of eq.~(\ref{rotation}).  Equivalently, the characteristic function
of $\sphi=\sphi^{(-)}\cup\sphi^{(+)}$ reads
\begin{equation}\label{chisphi}
  \chi(\sphi) =
  \Theta\big((-x^+ - \pi R\bar\Delta_\phi)(-x^- - \pi R\Delta_\phi)\big) +
  \Theta\big((x^+_R - \pi R\bar\Delta_\phi)(x^-_R - \pi R\Delta_\phi)\big) \;.
\end{equation}

The ensuing improved metric we propose is thus
\begin{align}
 \dif s^2 &= -\dif x^+\dif x^- 
 \left[1-\half \chi(\sphi)(1-\dot\rho)\right]
 \nonumber \\
 &+(\dif x^+)^2 \delta\big(x^{+}-\pi R \sgn(x^-)\bar\Delta(\xt)\big) \left[
   2\pi R \bar{a}(\xt) -\fourth(1-\dot\rho) |x^{-}-\pi R \sgn(x^+)\Delta(\xt)|
 \right]
 \nonumber \\
 &+(\dif x^-)^2 \delta(x^{-}-\pi R \sgn(x^+)\Delta(\xt)\big) \left[
   2\pi R a(\xt) -\fourth(1-\dot\rho) |x^{+}-\pi R \sgn(x^-)\bar\Delta(\xt)|
 \right]
 \nonumber \\
 &+ \dif r^2\left[1+2(\pi R)^2\chi(\sphi)\dot{\phi} \right]
 +\dif\theta^2\,r^2 \left[1+2(\pi R)^2\chi(\sphi)(\dot{\phi}
  +2r^2\ddot{\phi})\right] \;.
 \label{metrica}
\end{align}
It takes into account all the physics results following the analysis of the
improved eikonal model and the ACV effective action, and is self-consistent, in
the sense that the trajectory shifts caused by the shock-wave metric are felt by
the metric itself.

\begin{figure}[ht]
  \centering
  \includegraphics[width=0.47\linewidth]{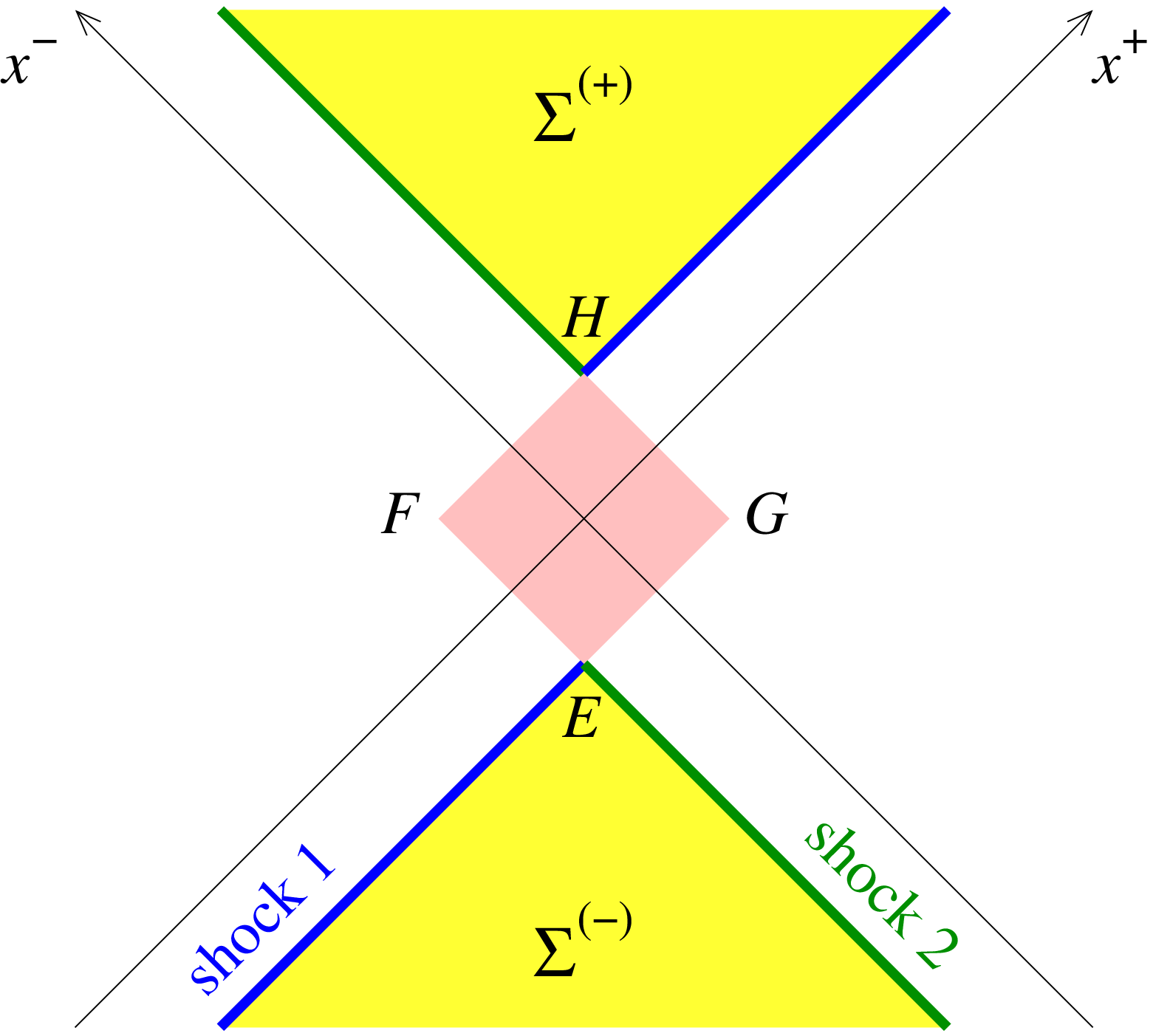}
  \Caption{Longitudinal section of the space-time (at the transverse coordinate
    of the center of mass) showing the position of the shock waves 1(blue) and 2
    (green) and the support $\sphi$ of the metric field $\Phi$ (yellow). The
    (pink) rectangle $EFHG$ actually corresponds to two different regions of
    space-time, which are accessed either from the left or from the right.  }
\label{f:wedgeSlice}
\end{figure}

A word of caution has to be spent concerning the form of the metric in
eq.~(\ref{metrica}) in the region between the forward and past light-wedges. In
fact, the set $|x^+|<\pi R\bar\Delta$, $|x^-|<\pi R\Delta$ enclosed by the
continuation of the shock waves --- represented by a rectangle $EFHG$ for each
longitudinal section at fixed $\xt$ as depicted in fig.~\ref{f:wedgeSlice} ---
actually represents two distinct physical regions in our coordinate system. Each
region is unambiguously described by continuing the metric either from the left
or from the right. For instance, by continuing the metric from the left across
the boundary $EFH$, we enter the region at the left of both shocks, where shock
1 travels from $E$ to $G$, collides with shock 2 in $G$, and after that shock 2
moves from $G$ to $H$ and goes on in the $x^-$ direction, as shown.  On the
other hand, either continuation leads to the same form of the metric inside the
future light-wedge $\sphi^{(+)}$.

We have also investigated here the scattering features of UV-sensitive solutions
of the model in the subcritical range $\lambda_s < b,r \lesssim R$, without
making a real attempt to restore the string degrees of freedom --- which are
nevertheless called for by the singular short-distance behaviour of the
rescattering current. This preliminary analysis leads to the suggestion that
such solutions may indeed carry the information from distances of order $R$ to
distances of order $\lambda_s$, but a dynamical analysis in the string-dominated
region is needed to find out about their fate.

%%%%%%%%%%%%%%%%%%%%%%%%%%%%%%%%%%%%%%%%%%%%%%%%%%%%%%%%%%%%%%%%%%%%%%%%%%%%%%%%
\section*{Acknowledgements}
%%%%%%%%%%%%%%%%%%%%%%%%%%%%%%%%%%%%%%%%%%%%%%%%%%%%%%%%%%%%%%%%%%%%%%%%%%%%%%%%

It is a pleasure to thank Gabriele Veneziano for various exchanges on the topics
presented here, and Domenico Seminara for useful discussions.

%%%%%%%%%%%%%%%%%%%%%%%%%%%%%%%%%%%%%%%%%%%%%%%%%%%%%%%%%%%%%%%%%%%%%%%%%%%%%%%%%
%%%%%%%%%%%%%%%%%%%%%%%%%%%%%%%%%%%%%%%%%%%%%%%%%%%%%%%%%%%%%%%%%%%%%%%%%%%%%%%%%
\appendix
\section*{Appendix}
%%%%%%%%%%%%%%%%%%%%%%%%%%%%%%%%%%%%%%%%%%%%%%%%%%%%%%%%%%%%%%%%%%%%%%%%%%%%%%%%%
%%%%%%%%%%%%%%%%%%%%%%%%%%%%%%%%%%%%%%%%%%%%%%%%%%%%%%%%%%%%%%%%%%%%%%%%%%%%%%%%%

%\noindent{\Large\bf Appendices}

%%%%%%%%%%%%%%%%%%%%%%%%%%%%%%%%%%%%%%%%%%%%%%%%%%%%%%%%%%%%%%%%%%%%%%%%%%%%%%%%%
\section{H-diagram insertions\label{a:hdi}}
%%%%%%%%%%%%%%%%%%%%%%%%%%%%%%%%%%%%%%%%%%%%%%%%%%%%%%%%%%%%%%%%%%%%%%%%%%%%%%%%%

We start noticing that the field-insertions we shall consider are basically
either {\it(a)} external (fig.~\ref{f:HdiagramInsertions}.a) or {\it(b)}
internal (fig.~\ref{f:HdiagramInsertions}.b,c). Here we shall concentrate on the
final external insertion kinematics which --- in the frame characterized by
initial particle's momenta $p_1+q,p_2$ and final momenta $k,k_2$ --- is able to
provide the full shift $2\pi R\Delta$ from past to future.

The field $\hti^{++}$ is coupled to the charge $\kappa(k^+ + q^+)(k^+)$ and thus
the insertion factor in front of the amplitude $\ui\amp$ is
\begin{equation}\label{htpp}
 \hti^{++}=\ui\kappa\int\frac{\dif q^+\dif q^-}{4\pi^2}
 \frac{k^+ \esp{-\frac{\ui}{2}(q^- x^+ + q^+ x^-)}}{
     \left[k^- + q^- -\frac{(\kt+\qt)^2-\ui\e}{k^+ + q^+}\right]
     \left(q^- - \frac{\qt^2-\ui\e}{q^+}\right) q^+}
 \frac{\dif^2\qt}{(2\pi)^2} \esp{\ui\qt\cdot\xt}
\end{equation}
where $k^+ + q^+$ and $k^+$ are both taken to the positive and larger than
$|q^+|$. The value of $k^- = \kt^2/k^+$ is fixed by the mass-shell and the
$q^-$-integration is done in the lower half-plane because $k^+ + q^+>0$ picking
up the pole at $q^-=\frac{(\kt+\qt)^2}{k^++q^+}-\frac{\kt^2}{k^+}$. The result is
\begin{align}
  \hti^{++} = \kappa\int\frac{\dif q^+}{4\pi} \esp{-\frac{\ui}{2} q^+ x^-} (k^+ + q^+)
  \frac{\dif^2\qt}{2\pi}
    &\left[
      \frac{\Theta(x^+)}{\left(\qt-\kt \frac{q^+}{k^+}\right)^2}
      \left(\esp{-\frac{\ui}{2}q^- x^+}-\Theta(q^+)\esp{-\frac{\ui}{2} \frac{\qt^2}{q+}x^+}
      \right) \right.\nonumber \\
 &\left. + \Theta(-x^+)\frac{\Theta(-q^+)\esp{-\ui\frac{\qt^2}{q^+}x^+}}{
        \left(\qt-\kt\frac{q^+}{k^+}\right)^2}
    \right] \esp{\ui\qt\cdot\xt} \label{htpp2}
\end{align}

Here we have kept for completeness the soft-emission contribution at $q^2=0$ also
which is however suppressed by the phase factor exponent
$\qt^2 x^+/q^+ \simeq x^- x^+ / r^2$ in the fast-emitting particle kinematics we
are using. The dominant contribution, due to the $(k+q)^2=0$ pole, has a small
phase factor $q^- x^+ \simeq x^+/(rbE) \ll 1$, and thus it
provides the $\Theta(x^+)$ result that we have anticipated in the text.%
\footnote{Note, however, that in the $x^+\to0^+$ limit the whole
  contribution~(\ref{htpp}) vanishes for $q^+>0$, because the $q^-$-integrand has
  both poles in the same half-plane.}

Note also that the $q$-translation in the $\qt^2$ denominator is due to the
scattering angle implied by the $\kt$ momentum transfer, which in a single-hit
process is small, $\tan(\theta/2) \simeq |\kt|/k^+ = \ord{1/(\sqrt{s}b)}$ and is
thus negligible. On the other hand, in the full scattering process we have that
$\sin(\theta/2) \simeq Q/\sqrt{s}$ is sizeable, and the shock-wave is
correspondingly rotated, as in eqs.~(\ref{hR},\ref{hppR}) of the text.

Therefore, for insertion on the eikonal lines in which $k^+\simeq p^+=\sqrt{s}$
and the amplitude factor is $2\pi G s\,\ui\As$, we recover the insertion factor
\begin{equation}\label{insFact}
  h_{--}=\frac{\kappa}{2}\,\hti^{++}=2\pi R(\ui\sqrt{s}-2\partial_-)
  \Theta(x^+)\delta(x^-) a_0(r)
\end{equation}
that we have repeatedly used in the text. Here $a_0$ is the leading profile
function of eq.~(\ref{a0}) and the possible rotation along the scattered beams
is understood.

In the case of the H-diagram, besides the insertion {\it(a)} on external lines
(already computed in the text) we have in principle two insertions, one of which
is on the internal (eikonal) propagators. We think that this one
(fig.~\ref{f:HdiagramInsertions}.b) is already included by the Lipatov's vertices,
which by definition count the external particles' emissions also. Therefore, in
order to avoid double-counting we only estimate here the insertion on the
rescattering graviton propagator of momentum $k$ in
fig.~\ref{f:HdiagramInsertions}.c, as emitted by Lipatov's vertices.

In order to do that, we compute the imaginary part of the $q$-insertion
amplitude, which has the form
\begin{align}
  2\Im\Delta h^{++}(x) = 2\pi G s R^2 \int\dif^2\kt\frac{\dif^2\qt}{(2\pi)^2}\;
  h^*(\kt)h(\kt+\qt)\frac{\dif k^+ \dif k^-}{8\pi^2}
  \frac{\dif q^+\dif q^-}{4\pi^2}
  \esp{-\frac{\ui}{2}(q^- x^+ + q^+ x^-)}\esp{\ui\qt\cdot\xt} \nonumber\\
  \times 2\pi\delta\left(k^- -\frac{\kt^2}{k^+}\right)
  \left(k^- + q^- - \frac{(\kt+\qt)^2-\ui\e}{k^+ + q^+}\right)^{-1}
  \left(q^- - \frac{\qt^2-\ui\e}{q^+}\right)^{-1} (q^+)^{-1} \;. \label{imDelta}
\end{align}

By going through the same steps as before, we perform the $q^-$-integration first
(by keeping the leading $(k+q)^2=0$ contribution only) and we have, in addition,
a weighted $k^+,\kt$-integration, as follows:
\begin{align}\label{didhpp}
  2\Im\Delta h^{++}(x) = \kappa\Theta(x^+)G s R^2
 &\int\dif^2\kt
  \frac{\dif^2\qt}{2\pi^2}\;h^*(\kt) h(\kt+\qt) \\ \nonumber
 &\times\int\frac{\dif k^+\dif q^+}{4\pi}\;\frac{k^+ + q^+}{2 k^+}
  \frac{\esp{\ui\qt\cdot\xt}}{\left(\qt-\kt\frac{q^+}{k^+}\right)^2}
  \esp{-\frac{\ui}{2} q^+ x^-} \;.
\end{align}

In the strong ordering region $\sqrt{s}\gg k^+ \gg |\kt|\sim 1/b,|q^+|$ we
estimate the effect of the (small) $q$-translation (or $\kt$-dependent rotation)
by redefining $\tilde{\qt}=\qt-\kt\, q^+/k^+$ and neglecting the translation in the
matrix element $h(\kt+\qt)$. The integral in $\dif^2\qt$ becomes
\begin{equation}\label{qtint}
  \int\frac{\dif^2\tilde{\qt}}{2\pi^2}\; \frac{\esp{\ui\tilde{\qt}\cdot\xt}}{\tilde{\qt}^2}
  \esp{\ui\kt\cdot\xt\, q^+/k^+} h^*(\kt)h(\kt+\tilde{\qt})
\end{equation}
by thus factorizing the exponential $\esp{\ui\kt\cdot\xt\,q^+/k^+}$ which
modifies the subsequent $k^+$ and $q^+$ integrations. By performing the latter
in the phase-space $-q^+ < k^+ < \infty$, we obtain in turn the factor
\begin{align}
  &\simeq \ui\partial_-\int_{|\kt|}^{\sqrt{s}}\frac{\dif k^+}{k^+}
  \left[\delta\left(x^- -2\frac{\kt\cdot\xt}{k^+}\right)-\frac1{2\pi}
  \frac{\esp{-\frac{k^+}{2}\left(\e-\ui(x^- - 2\kt\cdot\xt/k^+)\right)}}{
    \e-\ui\left(x^- -2\frac{\kt\cdot\xt}{k^+}\right)} \right] \nonumber \\
  &\simeq \log(\sqrt{s}b) \;\ui\partial_- \delta(x^-) \times
  \left[1+\ord{\frac1{\sqrt{s}b}}\right] \;.
\end{align}
We can see here the small shock-wave rotation
$\tan(\theta/2) \simeq |\kt|/k^+ = \ord{1/(\sqrt{s}b)} \simeq\esp{-y}$
which has been neglected in front of the logarithmic phase space. By then
collecting the various factors we obtain the result
\begin{equation}\label{imdelta}
  \Im\Delta h_{--} = \frac{\kappa}{2}\Im\Delta\hti^{++} = 2\pi R\Theta(x^+)
  \cdot 2\delta'(x^-) \cdot \pi R [G_0\Im a_H] \;,
\end{equation}
where we have defined
\begin{align}\label{G0def}
  [G_0\Im a_H](\xt) &\equiv \pi R^2 Y\int\dif^2 \xt'\;|\hti(\xt')|^2 G_0(\xt-\xt')
  \stackrel{r\gg b}{\simeq} a_0(r)\cdot\Im a_H(b) \;, \\
  \Im a_H &= \frac{2Y}{\pi}\Re a_H \;. \nonumber
\end{align}
This shift of the imaginary part and the corresponding one for the real part
add up to the total shift $2 a_H$ of both real and imaginary parts due to the
external line insertion
\begin{equation}\label{deltaAh}
  \left.\Delta(\ui\A_H)\right|_{\mathrm{ext}} = 2\pi R\Theta(x^+)
  \pi R a_H \left(\ui\sqrt{s}-4\partial_-\right)\delta(x^-)
\end{equation}
and thus occurs at large distances only with a relative minus sign, as
emphasized in the text.

%%%%%%%%%%%%%%%%%%%%%%%%%%%%%%%%%%%%%%%%%%%%%%%%%%%%%%%%%%%%%%%%%%%%%%%%%%%%%%%%

\end{document}